\documentclass[preprint,review,12pt]{elsarticle}

\journal{Solid State Sciences}

\usepackage{amssymb}
\usepackage{amsmath}

\usepackage{graphicx}
\usepackage{hyperref}
\usepackage{xspace}
\usepackage{subcaption}
\usepackage{pdfpages}

\newcommand{\ea}{~\textit{et~al.}\xspace}  
\newcommand{\md}[1]{\mathrm{#1}}           
\newcommand{\bm}[1]{\mathbf{#1}}           

\newcommand{\changed}[1]{\textcolor{black}{#1}} 

\begin{document}
\begin{frontmatter}

\title{
Data-Driven Prediction of NaCl-Type Entropy-Stabilized Oxide Compositions from First-Principles and Supervised Learning.
} 
\author[icmpe]{Sébastien Junier} 
\author[icmpe]{Céline Barreteau} 
\author[icmmo]{David Berardan} 
\author[icmmo]{Yann-Andrev Kerneur}
\author[link]{Jean-Claude Crivello} 

\affiliation[icmpe]{organization={Univ Paris Est Creteil, CNRS, ICMPE, UMR 7182},
            addressline={2 rue Henri Dunant}, 
            city={Thiais},
            postcode={94320}, 
            country={France}}
            
\affiliation[link]{organization={CNRS – Saint-Gobain – NIMS, IRL3629, Laboratory for Innovative Key Materials and Structures (LINK), National Institute for Materials Science},
            addressline={1-1 Namiki}, 
            city={Tsukuba, Ibaraki},
            postcode={305-0044}, 
            country={Japan}}
            
\affiliation[icmmo]{organization={Université Paris-Saclay, CNRS, Institut de chimie moléculaire et des matériaux d'Orsay},
            city={Orsay},
            postcode={91405}, 
            country={France}}
\begin{abstract}
Entropy-stabilized oxides (ESOs) open access to vast multicomponent compositional spaces, but identifying promising candidates remains challenging because of the large number of possible mixtures and the need to assess their stability against competing phases. 
In this work, we develop a high-throughput computational framework to screen equimolar quinary ESOs in the NaCl structure type by combining density functional theory (DFT), special quasirandom structures (SQS), convex-hull thermodynamics, and supervised machine learning. 
A consistent reference database of binary and ternary ordered oxides, including disordered phases such as all binary cation combinations in the NaCl-type oxide, is first constructed using GGA and meta-GGA calculations. 
Quinary disordered phases are then described by SQS supercells and used to train machine-learning models that predict the distance to the convex hull and the corresponding stabilization temperature over the full set of 4,368 possible equimolar quinary compositions generated from 16 cation species. 
Among the tested models, an optimized multilayer perceptron provides the best predictive performance, with a test error of about 4\,kJ$\cdot$mol$^{-1}$, while requiring explicit DFT calculations for only about 10\% of the quinary systems. 
The workflow successfully recovers known NaCl-type ESOs among the lowest predicted stabilization temperatures and provides a physically meaningful ranking of new candidate compositions. 
Comparison with experimental synthesis tests and computed decomposition paths further shows that the approach captures the main stability trends and the dominant competing phases, although absolute stabilization temperatures remain affected by systematic thermodynamic approximations. 
These results establish an efficient route for the data-driven exploration of multicomponent oxides and provide practical guidance for the experimental search for new ESOs.\\~\\
\end{abstract}

\begin{graphicalabstract}
\includegraphics{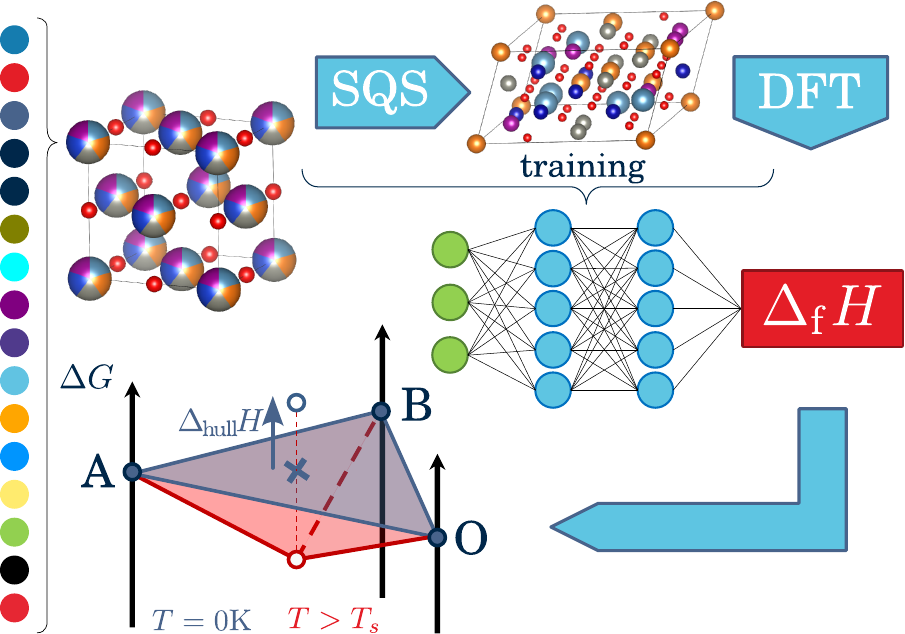}
\end{graphicalabstract}

\end{frontmatter}

\newpage
\section{Introduction}
    \label{sec:intro}
    
Designing entirely new multicomponent materials is a highly effective approach for developing inorganic compounds with unprecedented properties.
The discovery of high-entropy oxides (HEOs) by Rost\ea~\cite{Rost2015} has opened vast compositional spaces, by enabling the stabilization of single-phase structures through configurational entropy.
A HEO is a subclass of materials within the broader category of high entropy materials, composed of about five principal elements in near-equimolar proportions forming a solid solution, with additional oxygen anion in interstitial sites~\cite{Rost2015,Oses2020}. 
Unlike traditional oxides, which typically consist of one or two dominant elements, HEOs contain a mix of multiple metal cations on one single crystallographic site.
This characteristic may confer on these compounds a series of intriguing properties, such as a colossal dielectric constant~\cite{Berardan2016}, particular magnetic behaviours~\cite{Zhang2019,Sun2021} and more interesting, the possibility of multifunctional and tunable properties as a function of the chemical composition~\cite{Kumar2023,Jiang2018}.
The number of potential applications is therefore considerable including photovoltaics~\cite{Kumbhakar2023}, optics~\cite{Songbo2023}, catalysts~\cite{Chen2018} \changed{or photocatalysis~\cite{shan_high-entropy_2025}}, energy storage (e.g. Li-ion batteries), due to their high ionic conductivity~\cite{Hou2024}\changed{, thermal barrier coating~\cite{vakilifard_high_2024}, or fuel cells electrodes and electrolytes~\cite{gu_brief_2026}}.

Entropy-stabilized oxides (ESOs), the focus of this work, are a class of high-entropy oxides that are metastable at room temperature but can become stable at elevated temperature due to configurational entropy~\cite{Dragoe2019}. 
Although they can be retained by rapid quenching, identifying new stable ESOs remains challenging because of the vast chemical space accessible to multicomponent oxides. 
In principle, a given ESO composition may become stable above its characteristic stabilization temperature, $T_\mathrm{stab}$, if competing phases are neglected. 
In practice, however, this temperature must remain experimentally accessible and below the melting point. 
A brute-force experimental search would therefore require testing compositions one by one and checking whether they form a stable single-phase solid solution, which is highly time-consuming. 
Moreover, no general design rule has yet been established for selecting the most favorable set of elements, and recent studies have shown that mixing five ordered oxides with the same prototype is not necessarily the most effective strategy for stabilizing an ESO in that structure, as it mostly leads to thermodynamically stable HEOs~\cite{Rost2015,Kumar2023,Pitike2020}.

In this work, we develop a high-throughput computational workflow combining density functional theory (DFT) with special quasirandom structures (SQS) and supervised machine learning to screen equimolar quinary ESOs in the rocksalt structure with $n=5$ elements in addition to oxygen. 
Focusing exclusively on this single-phase NaCl prototype, the first ESO experimentally observed~\cite{Rost2015}, we aim to evaluate the thermodynamic stability of $\binom{16}{5}=4{,}368$ equimolar quinary ESO systems formed by mixing $N=16$ elements (Ca, Cu, Co, Fe, Hf, In, Mg, Mn, Ni, Sn, Sr, Ti, V, Y, Zn, Zr) on the cationic sublattice, with oxygen fixed on the anionic sites. 
The elements were selected according to several criteria: abundance, low toxicity, and broad chemical diversity, prioritizing transition metals, alkaline-earth metals, and $s^2p^x$ metals.
Here and throughout, "quinary system" refers to $n=5$ cations on the $4a$ site, plus oxygen on $4b$ of $Fm\bar{3}m$ space group, described as $(\{M_i\})$O, with $y_i=\frac1n$ the occupation fraction of $M_i$ for $i=\{1,\dots,n\}$ on $4a$.
By predicting formation enthalpies of HEO relative to the convex hull ($\Delta_\mathrm{hull}H$) and stabilization temperatures ($T_\mathrm{stab}$), we identify the most promising candidates for single-phase ESOs, validated against known benchmarks like (Co,Cu,Mg,Ni,Zn)O.

The paper is organized as follows. 
Section~\ref{sec:methods} presents the DFT--SQS workflow, the construction of the reference convex-hull database, and the supervised learning strategy used to identify promising NaCl-type ESO candidates. 
Section~\ref{sec:results} then reports the main results, including the validation of the reference database and convex-hull construction, the identification of promising quinary ESO candidates, and the comparison of the predictions with experimental synthesis trials and finite-temperature decomposition paths.

\section{Computational methods}
\label{sec:methods}

In this study, we adopt a high-throughput computational workflow grounded on fundamental thermodynamic principles to predict new quinaries ESO of equimolar composition $(\{M_i\})$O within a given crystal-structure prototype, exemplified here by the NaCl-type structure. 
The phase-stability assessment relies on a custom convex-hull construction in the six-dimensional composition space $\bm{x} = \{x_{M_1}, x_{M_2}, x_{M_3}, x_{M_4}, x_{M_5}, x_\mathrm{O}\}$, built from a database of ordered oxides and disordered oxides $(\{M_i\})$O with $2\le n\le5$.
Within this space, the convex hull is defined as the minimal convex set containing all known phases, such that compositions lying on the hull correspond to stable single phases and those above the hull decompose into the nearest stable phases.

\subsection{Disorder Modelling via Special Quasirandom Structure}
\label{subsec:SQS}

\begin{figure}
        \begin{subfigure}{0.47\textwidth}
            \centering
            \caption{}
            \includegraphics[width=\textwidth]{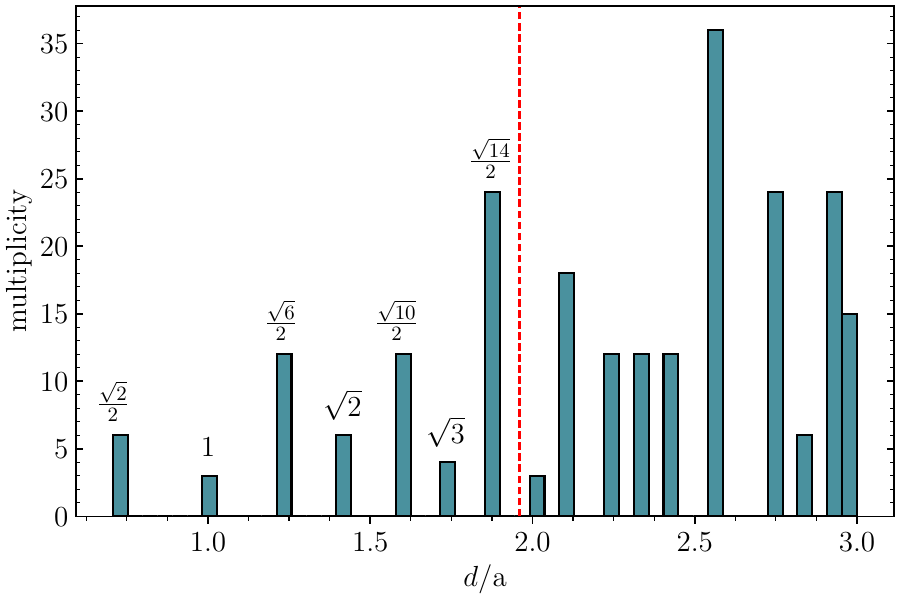}
        \label{fig:clusters-disctibution}
    \end{subfigure}\hfill
    \begin{subfigure}{0.47\textwidth}
            \centering
            \caption{}
            \includegraphics[width=\textwidth]{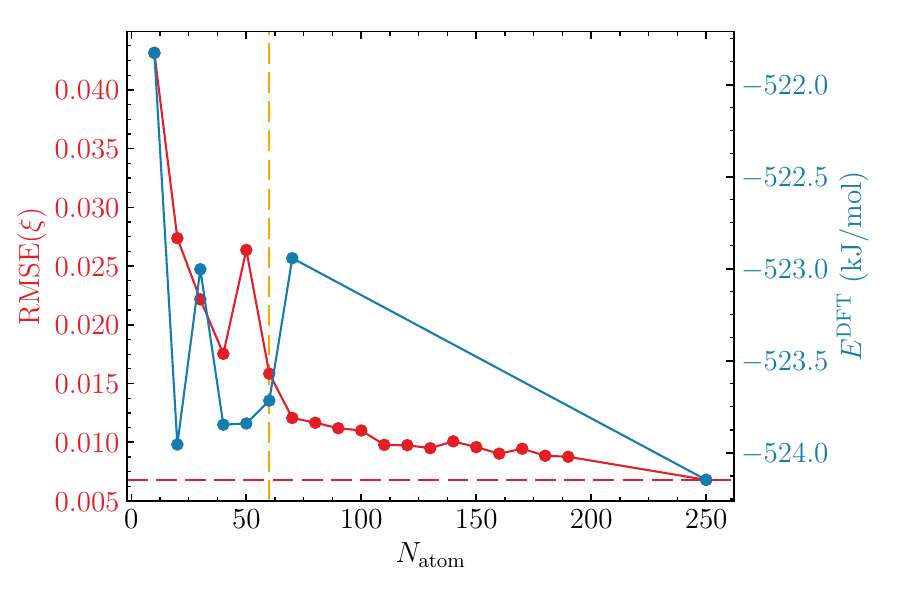}
        \label{fig:rmse-quinary}
    \end{subfigure}
    \caption{(a) Pair clusters distribution in NaCl structure. All clusters under the red lines are selected. the distribution is plot in function of the $d$/a where $d$ is the diameter of cluster and a the parameter of conventional cell. 
    (b) RMSE of clusters correlation function and DFT total energy of a quinary mixture, (Ca$_{0,2}$Cu$_{0,2}$Mg$_{0,2}$Sn$_{0,2}$Zn$_{0,2})$O for this example, as a function of atoms number in SQS supercell.}
    \label{fig:sqs}
    \end{figure}

To accurately model atomic disorder in equimolar solid solutions, we employ the SQS method~\cite{Zunger1990}, implemented via the Alloy Theoretic Automated Toolkit (ATAT)~\cite{Walle2002.1,Walle2002.2,Walle2009,Walle2013}.
SQS generates finite supercells whose multi-site correlation functions statistically reproduce those of a perfectly random solid solution up to a chosen cluster expansion order. 
For the NaCl-type rocksalt structure, pair clusters up to the 7th nearest neighbor are included; triplet clusters are omitted as preliminary tests on binary ($M_1,M_2$)O mixtures ($n=2$) showed no improvement in correlation convergence for this structure.
The same cluster set is used consistently across all mixing degrees (binary, ternary, and quinary) for mutual comparability.
Supercell sizes were optimized by monitoring both the pair-correlation root-mean-square error (RMSE) and the DFT total energy as a function of cell size, as shown in Fig.~\ref{fig:rmse-quinary}.
Convergence of the correlation RMSE toward zero is required for a supercell to be considered representative of the random mixture and the DFT energy convergence provides an estimate of the residual SQS error relative to the infinite-size limit. 
The chosen SQS sizes are detailed into the Result~\ref{subsec:result-SQS} section.

\subsection{First-Principles Calculations}
\label{subsec:DFT}

All \textit{ab initio} calculations done in this work employed the DFT~\cite{Hohenberg1964,Kohn1965} as implemented in VASP code~\cite{Kresse1993,Kresse1994,Kresse1996}. 
Wave functions are expanded in a plane-wave basis with a kinetic energy cutoff of 600\,eV, using projector-augmented wave (PAW) pseudopotentials~\cite{Blochl1994,Kresse1999}.
Exchange-correlation effects are treated at two levels of approximation.
The GGA functional of Perdew, Burke, and Ernzerhof (PBE)~\cite{Perdew1996} is used for all large-scale structural relaxations and database construction, providing a consistent and computationally affordable baseline. 
The meta-GGA r$^2$SCAN functional~\cite{Bartok2019,Furness2022} is applied selectively, for benchmark compounds and stabilization temperature validation and to improve accuracy, given its superior treatment of electron localization in transition metal oxides and its systematically lower convex-hull energies ($\sim$10--20\,kJ/mol relative to PBE) according to our tests on oxides.

Spin polarization is applied systematically to all ordered reference oxides containing magnetic elements (Co, Fe, Mn, Ni), initialized with ferromagnetic configurations. 
Magnetic ordering is not manually optimized, as doing so would be incompatible with the large number of compounds in the database; this introduces residual uncertainties for strongly correlated systems, discussed in Section~\ref{subsec:result-hull}. 
For SQS supercells, an initial spin-polarized calculation is always performed; if the converged total magnetic moment is $< 0.1\,\mu_B$, a non-spin-polarized refinement is subsequently carried out to reduce computational cost.
Brillouin-zone integration uses Monkhorst-Pack $k$-meshes optimized for convergence to $< 1$\,meV/atom: a $15\times15\times15$ grid for the primitive 2-atom rocksalt unit cell, and meshes such as $5\times4\times3$ for 60-atom quinary SQS supercells. 
Full details of $k$-mesh convergence tests are provided in the Supplementary Material~A.

The formation enthalpy of oxide at $\bm{x}_\varphi$ composition is computed as:
\begin{equation}
    \Delta_\mathrm{f}H(\bm{x}_\varphi) =
    E^\mathrm{DFT}_\varphi - \sum_i x_i\,E^\mathrm{SER}_i,
    \label{eq:formation-enthalpy}
\end{equation}
where $E^\mathrm{DFT}_\varphi$ is the DFT total energy of $\varphi$, $E^\mathrm{SER}_i$ the DFT energy of pure element $i$ in its Standard Element Reference (SER) state, and $x_i$ its atomic composition. 

Zero-point energy (ZPE) corrections are computed from harmonic $s$~phonons calculations using Phonopy code ~\cite{Togo2015,Togo2023}, $E_\md{ZPE}=\frac12\sum_{\bm{q},s}\hbar\omega(\bm{q},s)$ in $\bm{q}$ reciprocal space, performed on all ordered oxides in the NaCl structure to support the ZPE linear approximation of Eq.~\eqref{eq:zpe-mix}.
The ZPE correction for SQS mixtures is not computed directly from phonon calculations, as the large supercell sizes make this prohibitively expensive. 
Instead, it is approximated as a linear combination of the phonon-calculated ZPEs of the ordered end-member oxides:
\begin{equation}
    E_\mathrm{ZPE}^\mathrm{SQS}\!\left[(\{M_i\})\mathrm{O}\right]
    \approx \sum_i y_i\,E_\mathrm{ZPE}(M_i\mathrm{O}),
    \label{eq:zpe-mix}
\end{equation}
where $M_i\mathrm{O}$ defines an ordered oxide and $(\{M_i\})\mathrm{O}$ a solid solution with a mixture of several element $M_i$ with a proportion $y_i$.   

This approximation implies a negligible ZPE contribution to the mixing enthalpy, which was validated on the binary (Mg,Sr)O system: a full phonon calculation on the 32-atom SQS supercell yields a ZPE correction to the mixing enthalpy of $<1\%$ ($\Delta_\mathrm{mix}H$ changes from 49.11 to 48.64\,kJ/mol), confirming the validity of Eq.~\eqref{eq:zpe-mix} for this class of oxides.

\subsection{Thermodynamic approximations for the finite-temperature convex hull}

Phase stability is assessed via a custom convex-hull construction in composition--energy space, built from the reference database. 
For any multicomponent oxide $\varphi$ at composition $\bm{x}_\varphi$, the distance to the hull is:
\begin{equation}
    \Delta_\mathrm{hull}H(\bm{x}_\varphi) =
    \Delta_\mathrm{f}H(\bm{x}_\varphi) - E^\mathrm{GS}(\bm{x}_\varphi),
    \label{eq:dhull}
\end{equation}
where $E^\mathrm{GS}(\bm{x}_\varphi)$ is the ground-state hull energy at that $\bm{x}_\varphi$ composition. 
If $\Delta_\mathrm{hull}H > 0$, the compound is unstable at 0\,K and decomposes into the adjacent hull phases; 
if $\Delta_\mathrm{hull}H = 0$, it lies on the hull and is thermodynamically stable.

First, a comprehensive database of ordered reference oxides (binaries and ternaries), extracted from experimentally observed compounds containing the 16 selected cations~\cite{PCD}, is constructed via first-principles calculations to build the whole convex hull $E^\mathrm{GS}(\bm{x})$.
Then, since DFT calculations are performed at 0\,K, finite-temperature effects for disordered oxides are considered via configuration entropy contribution in the Gibbs free energy:
\begin{equation}
    \Delta_\mathrm{f}G = \Delta_\mathrm{f}H - T\,\Delta S_\mathrm{conf}.
    \label{eq:DG}
\end{equation}
For HEOs with five or more cations in near-equimolar proportions, configurational entropy dominates and is approximated by Boltzmann's expression:
\begin{equation}
    \Delta S_\mathrm{conf} = -R \sum_{i=1}^{n} y_i \ln( y_i),
    \label{eq:DS}
\end{equation}
with $R = 8.314$\,J$\cdot$K$^{-1}\cdot$mol$^{-1}$.
This does not yield $\Delta S_\mathrm{conf} \approx 1.61\,R$ as in a classical fully equimolar quinary mixture; instead, this value is effectively halved as compared to per formula unit because oxygen occupies one ordered sublattice while the $n$ cations share the other in NaCl-type $(\{M_i\})$O ESOs.

Then, to determine $T_\mathrm{stab}$, we developed an iterative convex-hull algorithm based on the Quickhull method~\cite{Barber1996}. 
The algorithm constructs the ground-state hull from $\Delta_\mathrm{f}H$ in the $\{\bm{x}_i,\Delta_\mathrm{f}H\}$ composition--energy space, where $\bm{x}_i$ denotes the atomic fraction of each of the 16 cations and oxygen. 
Compound stability is quantified by 
$\Delta _\mathrm{hull}H(\varphi)$~(Eq.~\eqref{eq:dhull}):
stable phases satisfy $\Delta_\mathrm{hull}H=0$ and define the hull facets, whereas metastable compounds have $\Delta_\mathrm{hull}H>0$ and decompose into the nearest stable phases according to the lever rule.

At finite temperature, $\Delta_\mathrm{f}G(T)$ replaces $\Delta_\mathrm{f}H$ (Eq.~\eqref{eq:DG}), leading to a temperature-dependent hull energy $E^\mathrm{GS}(\bm{x},T)$. 
Because the configurational entropy of an equimolar quinary lowers its free energy faster than the enthalpy-driven hull of ordered oxides, a stabilization temperature $T_\mathrm{stab}$ can be defined as the point where the ESO free energy first intersects the hull:
\begin{equation}
    \Delta_\mathrm{hull}G(\bm{x}_\varphi,T_\mathrm{stab}) =
    \Delta_\mathrm{f}G(\bm{x}_\varphi,T_\mathrm{stab})
    - E^\mathrm{GS}(\bm{x}_\varphi,T_\mathrm{stab}) = 0.
    \label{eq:Tstab}
\end{equation}
Substituting Eq.~\eqref{eq:DG} gives the iterative update:
\begin{equation}
    T_{i+1} =
    \frac{\Delta_\mathrm{f}H(\bm{x}_\varphi) - E^\mathrm{GS}(\bm{x}_\varphi,T_i)}
         {\Delta S_\mathrm{conf}},
    \label{eq:iter}
\end{equation}
starting from $T_0=0$\,K and repeated until $|T_{i+1}-T_i|<1$\,K. 
The Python implementation operates efficiently in up to 17-dimensional composition space, covering all sub-chemical systems of the 16-cation library. 
The code is available in the Code availability section.
At the end, the $(\{M_i\})$O compositions with the lowest $T_\mathrm{stab}$ are prioritized as the most promising ESO synthesis targets.

\subsection{Machine learning models}
\label{subsec:ML}
To screen $\binom{16}{5}=4{,}368$ equimolar quinary systems affordably, supervised machine learning has been use to predict $\Delta_\mathrm{hull}H$ from any $\bm{x}_\varphi$ composition. 
Training database was built from the consistent DFT $\Delta_\mathrm{f}H$ for ordered rocksalt, binaries, ternaries, and quinary SQS across the 16 cations, that have been done during this work, under same condition.

In addition to chemical composition expressed using One-Hot-Encoding based on cation fractions $\{y_i\}$ with $\sum y_i=1$, several chemical features have been considered, as the mean and the deviation for six selected properties $p$ (e.g., radius $r_\text{at}$, electronegativity $\chi$ as shown in Table~\ref{tab:descriptor-symbols}), included from Mendeleev library~\cite{Mentel2021}:
\begin{equation}
\label{eq:rmse-des}
\overline{p} = \sum_i y_i p_i, \quad \delta_p = \sqrt{\sum_i y_i (p_i - \overline{p})^2}.
\end{equation}

Performance have been evaluated using the classical score functions (MAE, RMSE, $\chi^2$), via Scikit-learn~\cite{scikit-learn} where three models have been benchmarked (Linear model, Random Forest, multi-perception MLP) using 4-fold cross-validation (CV) on quinaries for hyperparams optimizations.

\begin{table}[ht]
\centering
\caption{Chemical descriptors for ML models of rocksalt ESO stability.}
\label{tab:descriptor-symbols}
\begin{tabular}{ll}
\hline
\textbf{Symbol} & \textbf{Property} \\
\hline
$y_i$ & elemental fraction $i$ ($i=1,\ldots,16$) \\
$Z_\text{at}$ & Atomic number \\
$r_\text{at}$ & Metallic radius \\
$col$ & Periodic table column \\
$\rho_\text{at}$ & Atomic density \\
$\chi$ & Pauling electronegativity \\
$e^-_\text{val}$ & Valence electrons \\
\hline
\end{tabular}
\end{table}

\section{Results and discussion}
    \label{sec:results}

    \subsection{Reference Database Construction}
To determine the stability of hypothetical entropy-stabilized oxides (ESOs), their energies must be benchmarked against known stable ordered phases. 
We thus built a consistent database of ground-state formation energies $\Delta_\mathrm{f} H$ as a function of composition for simple ordered oxides from the 16 selected elements. 
All calculations used identical and consistent DFT conditions (PBE GGA and r$^2$SCAN meta-GGA functionals, with ZPE corrections from phonons).

Binary $M_{1}$O$_z$ ordered structures were selected from ambient stable pure elements and from frequent prototypes (at least five literature reports, sourced from the Pearson Crystal Database), including not only NaCl-type oxides with oxygen in octahedral coordination, but also many other structures, such as corundum Al$_2$O$_3$, rutile TiO$_2$, binary spinel Fe$_3$O$_4$, and baddeleyite ZrO$_2$. 
Together, these structures cover a broad range of oxygen coordination environments, such as triangular, tetrahedral, and octahedral sites.
As shown in Supplementary Materials B1 to B16, the binary results are consistent across all systems: 
all convex hulls of $\Delta_\mathrm{f}H$ as a function of oxygen content identify the known 0~K stable phases. 
Overall, the calculated ground states agree well with experiment, both qualitatively, since all experimentally observed phases lie on or very close to the hull, and quantitatively, with r$^2$SCAN errors below 4\% relative to experimental data. 
In addition, the hulls obtained with r$^2$SCAN are typically 10--20~kJ/mol lower than those obtained with PBE, indicating that the meta-GGA functional provides a much better reproduction of formation enthalpies.
The vibrational contribution is negligible, as the ZPE correction modifies the hulls by less than 2~kJ/mol. 
For magnetic systems (Mn/Co/Fe/Ni--O), larger discrepancies arise from the automated, non-optimized treatment of magnetic ordering. 
Nevertheless, r$^2$SCAN correctly stabilizes key phases such as Fe$_3$O$_4$, NiO, and MnO, for which the magnetic ordering was enforced to match the experimental one rather than being allowed to converge automatically during DFT relaxation.
Overall, the binary hulls remain suitable for relative ESO screening, although reduced accuracy should be expected for magnetic compounds.

For ordered ternary compounds $M1_{y1}M2_{y2}$O$_z$, approximately 30 known and reported crystal structures were selected, mainly from databases such as the Pearson Crystal Database, AFLOWlib, and the Materials Project. 
These structures are largely grouped into four oxide families: perovskites (CaTiO$_3$, $cP5$), spinels (MgAl$_2$O$_4$, $cF56$), rhombohedral $E2_2$ phases (FeTiO$_3$, $hR10$), and layered $tI14$ structures (K$_2$NiF$_4$), thereby covering a broad range of chemical environments.
All calculations were performed systematically under the same conditions as for the binaries. 
As expected, the known phases are thermodynamically stable at 0~K according to the ground-state analysis, with only rare exceptions where binary decomposition is observed. 
Again, the meta-GGA r$^2$SCAN provides better results, whereas many systems are predicted to be more stable in PBE functional than expected for some ternary phases. 
Full details, including crystal structure figures and tables of relative energies of ternary compounds are provided in the Supplementary Material~C. 
This uniform reference set enables reliable thermodynamic screening of ESOs.
   
    \subsection{SQS generation to described disordered $(\{M_i\})$O NaCl phase}\label{subsec:result-SQS}
   
The choice of cluster types is first optimized on the binary $(M_1,M_2)$O rocksalt mixture. 
Once SQS calculations including the first seven neighbour pairs are performed, a 32-atom supercell is obtained whose pair correlation functions exactly match those of an ideal random solution. 
The distribution of pair clusters in the NaCl structure (only one sublattice of atoms mixed as in FCC structure), and the subset retained for SQS generation, is reported in Fig.~\ref{fig:clusters-disctibution}. 
Adding higher-order clusters (triplets or beyond) either does not change the resulting supercell or degrades the quality of lower-order pair correlations; we therefore restrict the SQS construction to the first seven pair shells.
        
\begin{figure*}
        \centering
        \begin{subfigure}{0.45\textwidth}
            \centering
            \includegraphics[height=.2\textheight]{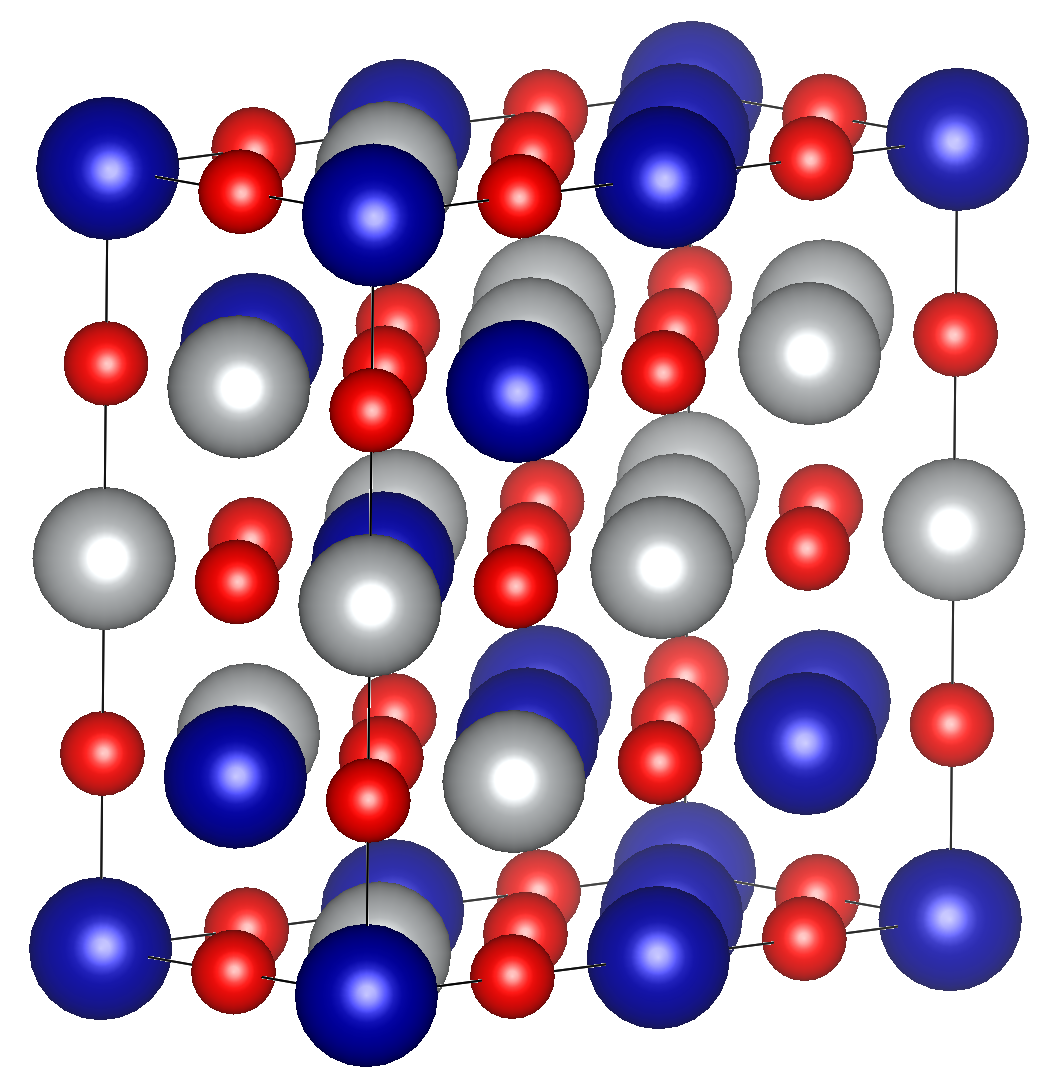}
            \caption{\centering}
            \label{subfig:sqs-binary}
        \end{subfigure}\hfill
        \begin{subfigure}{0.45\textwidth}
            \centering
            \includegraphics[height=.3\textheight]{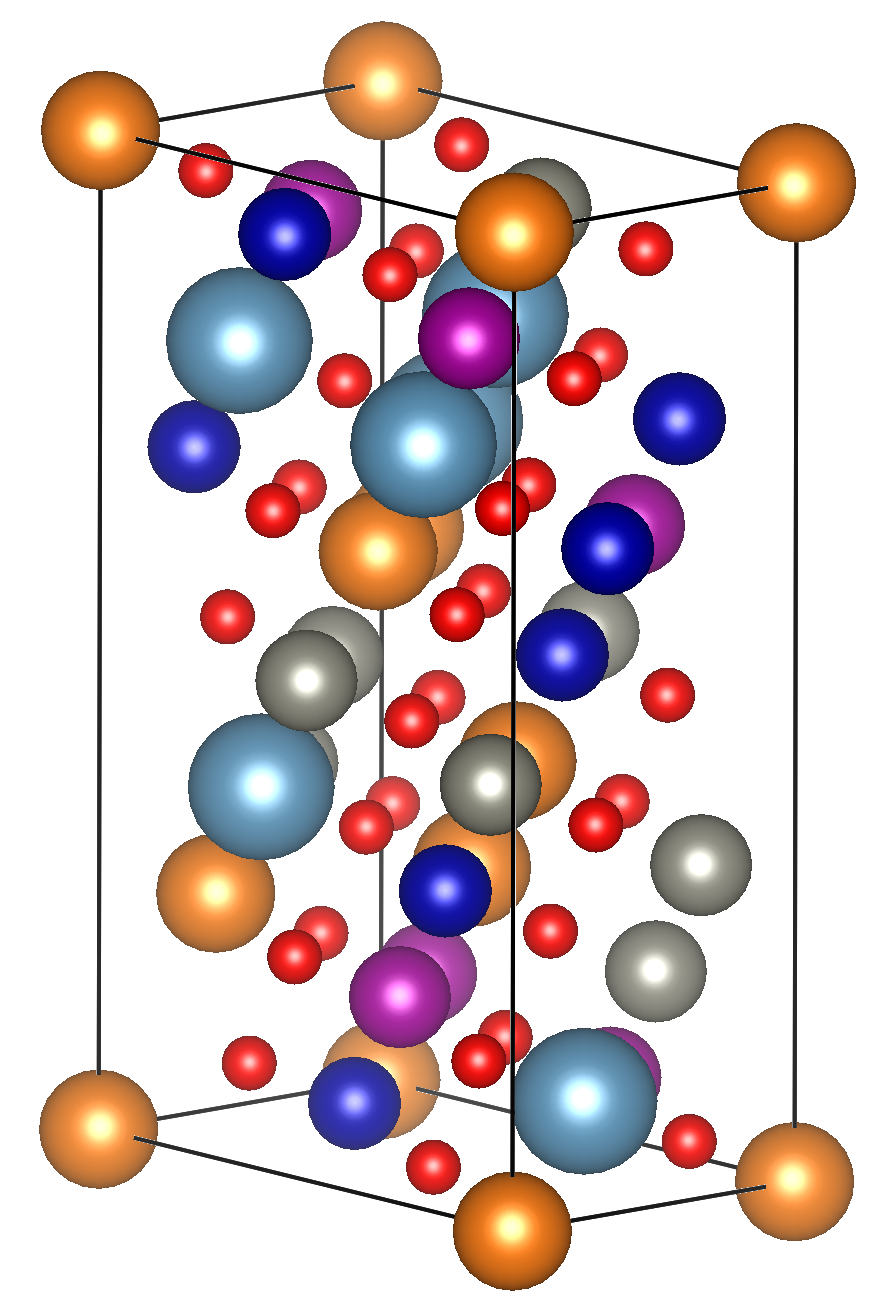}
            \caption{\centering}
            \label{subfig:sqs-quinary}
        \end{subfigure}
        \caption{Representation of equimolar super-cell generated by SQS for $(M_{1}M_{2})$O binary (\subref{subfig:sqs-binary}) and $(M_{1}M_{2}M_{3}M_{4}M_{5})$O quinary (\subref{subfig:sqs-quinary}) mixtures in a NaCl structure. Its contain 32 and 60 atoms respectively. Oxygen is red and the atom of mixing are in the other color.} 
    \end{figure*}

Using a similar strategy, the size of quinary SQS supercells is optimized by monitoring both the RMSE of correlation functions and the convergence of DFT total energies $E_\text{DFT}$ on a set of benchmark compositions. 
For equimolar quinary $y_i=\frac15$ of $(M_{1}M_{2}M_{3}M_{4}M_{5})$O systems, RMSE values rapidly decrease with increasing supercell size and become negligibly small above 60 atoms, while $E_\text{DFT}$ fluctuations remain below 1~kJ/mol per mole of atoms over 60 atoms~(Fig.~\ref{fig:rmse-quinary}). 
We therefore select 60-atom supercells ($5\times6=30$ cations and 30 oxygen atoms), corresponding to exactly 6~mixed $M_i$ cations per species. 
This choice provides a good compromise between an accurate description of chemical disorder and acceptable computational cost for the large set of 416 among the 4,368 quinary systems calculated by DFT in order to build the training database (Fig.~\ref{subfig:sqs-quinary}).
  
Zero-point energy (ZPE) corrections are evaluated selectively. 
Ordered oxides exhibit ZPE up to 5\% of $|\Delta_\mathrm{f}H|$. 
For solid solutions, the rule-of-mixtures approximation (Eq.~\eqref{eq:zpe-mix}) is validated on binary mixing (Mg,Sr)O, yielding mixing enthalpy changes $<1\%$ (Table~\ref{tab:zpe-MgSrO}) despite minor imaginary frequencies from SQS residuals. 
Since mixing enthalpies agree within 10\% with CALPHAD-assessed values, confirming the reliability of our approach, 
the ZPE value obtained from ideal mixing of binaries value is thus applied to ground-state phases and estimated for metastable quinaries.

    \begin{table}[h]
        \centering
        \begin{tabular}{ll}\hline
             \textbf{Calculation type} & ${\Delta_\md{mix}H}$ \textbf{(kJ/mol)}   \\\hline             
             DFT without ZPE     & 49.11                   \\
             DFT with ZPE        & 48.64                   \\
             Calphad assessment \cite{Kemp1994}  & 53.20                   \\\hline
        \end{tabular}
        \caption{Mixing enthalpy of (Mg,Sr)O in NaCl structure with and without ZPE calculation, in comparison with the work of van der Kemp\ea\cite{Kemp1994}. All enthalpy are given for one atomic mole.   }
        \label{tab:zpe-MgSrO}
    \end{table}

Another confirmation of the quality of the SQS cells is provided by the estimation of $E_\text{DFT}$ and the corresponding $T_\mathrm{stab}$ for the two phases observed experimentally, (Co,Cu,Mg,Ni,Zn)O and (Co,Fe,Mg,Mn,Ni)O. 
From the GGA-DFT calculations, the respective $T_\mathrm{stab}$ values are 2,745\,K and 3,177\,K. 
These values remain far from the experimental values of 1,100\,K~\cite{Rost2015} and 1,250\,K~\cite{Pu2023}, but additional meta-GGA calculations correct them to 1,249\,K and 2,567\,K, respectively, bringing them into better agreement with experiment and emphasizing that the meta-GGA treatment remains essential to deal with delocalised electrons from oxygen.

According to the present analysis, the selected SQS sizes (details and files are given in Supplementary Material~A) are finally chosen as follows.
\begin{itemize}
    \item 32 atoms for binary $(M_1,M_2)$O mixtures (8 cations per species
          + 16~O), with a residual energy uncertainty of $\sim$2\,kJ/mol;
    \item 48 atoms for ternary $(M_1,M_2,M_3)$O mixtures (8 cations per species
          + 24~O), with $\sim$1\,kJ/mol uncertainty;
    \item 60 atoms for equimolar quinary $(M_1,\dots,M_5)$O mixtures
          (6 cations per species + 30~O), with $\sim$1\,kJ/mol
          uncertainty.
\end{itemize}
    
\subsection[Prediction of deviation from convex-hull by supervised learning]%
           {Prediction of $\Delta_\mathrm{hull}H$ by supervised learning}
\label{subsec:result-hull}

From the 60-atom SQS description of quinary $(M_1,\dots,M_5)$O mixtures, DFT calculations were performed on 416 configurations, selected from a homogeneous distribution of elements as shown in Supplementary Material~D. 
These data constitute the training database for the machine-learning component, at a total cost of approximately one million core-hours for about 420 quinary systems for the only PBE relaxation steps.

\begin{figure}[ht]
    \centering
    \includegraphics[width=.8\linewidth]{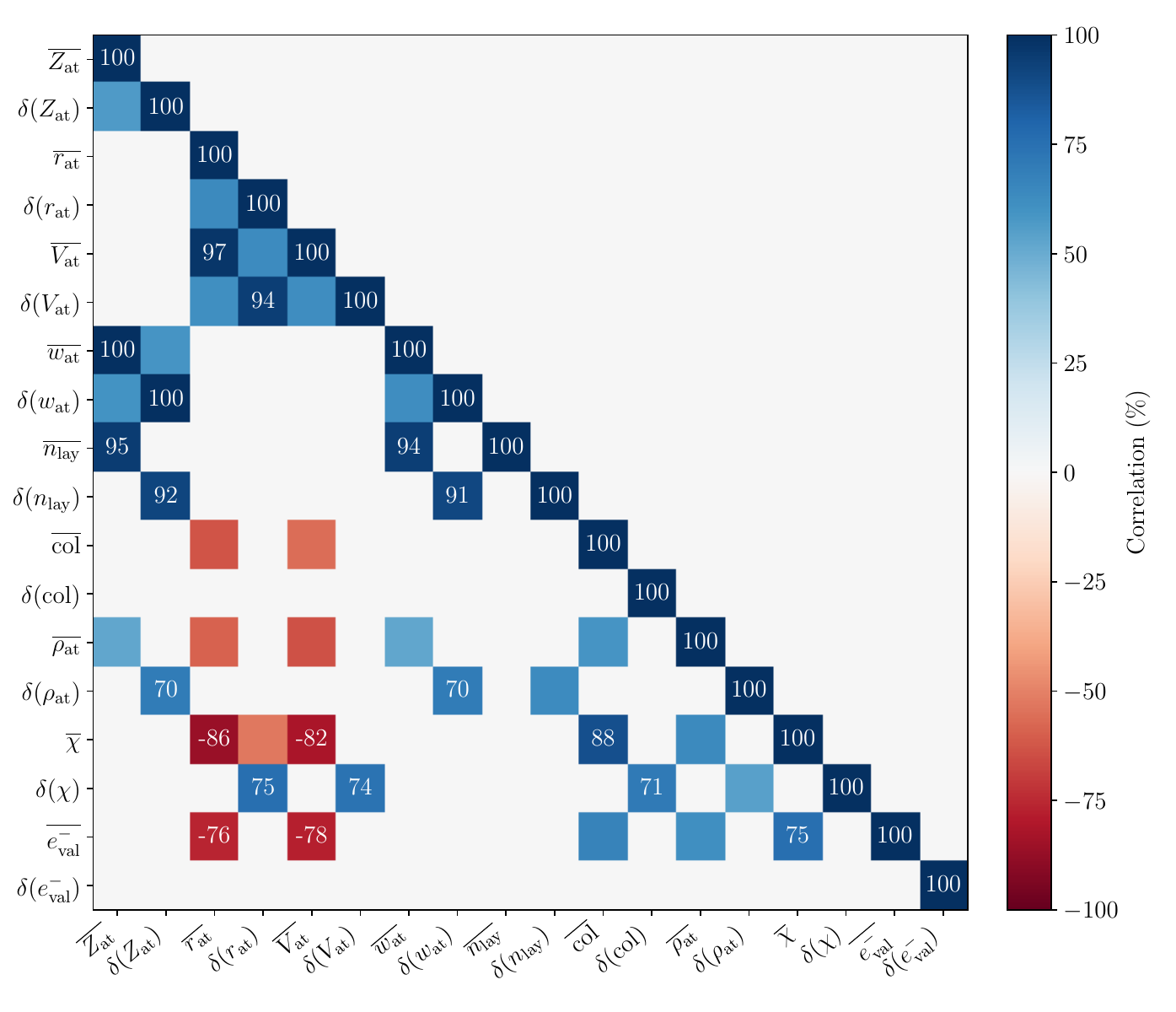}
    \caption{Correlation matrix of descriptors. 
    Correlations below 50\% are hidden; values above 70\% are annotated.}
    \label{fig:corr-matrix}
\end{figure}

The descriptors correlation matrix (Fig.~\ref{fig:corr-matrix}) reveals near-total correlations ($\approx$100\%) between atomic mass and atomic number, atomic volume and atomic radius, and number of electron shells and atomic mass. 
These redundant descriptors are removed, retaining seven features (Table~\ref{tab:descriptor-symbols}): cation fractions $y_i$, atomic number, atomic density, electronegativity, atomic radius, periodic table column, and number of valence electrons. 
Residual cross-correlations (70--80\%) between electronegativity and radius/valence electrons are retained as they do not follow strict functional relationships and introduce no significant bias.

Then, several supervised ML models were systematically benchmarked to predict $\Delta_\mathrm{hull}H$ for equimolar quinary NaCl-type oxides directly from chemical composition and physical descriptors. 
Three model families were evaluated: linear regression (LR), random forest (RF), and multilayer perceptron (MLP) neural networks. 
To ensure robust quinary prediction, a manual 4-fold cross-validation isolates quinary test sets: 
each fold trains on all 16 ordered $M_1$O unaries, 120 $(M_1,M_2)$O binary, and 27 $(M_1,M_2,M_3)$O ternary rocksalt systems, plus 312 of the 416 DFT-calculated $(M_1,\dots,M_5)$O quinary SQS supercells, reserving 104 quinaries for testing ($\sim$18\% of the total dataset).

Table~\ref{tab:RMSE} summarizes the training and test errors for each model using the full descriptor set. 
The LR model yields an acceptable test RMSE of $6.09 \pm 0.21$\,kJ/mol, already outperforming a naive binary combination linear rule, but the learning curve has not converged, indicating that more data would be needed to increase the prediction quality (Learning curves are given in Supplementary Material~E). 
The RF achieves a substantially lower test RMSE ($4.86 \pm 0.29$\,kJ/mol), with convergence confirmed by the learning curve; notably, using only composition fractions $y_i$ without additional physicochemical descriptors already yields $5.14 \pm 0.43$\,kJ/mol, highlighting the strong compositional signal. 
The MLP neural network with optimized hyperparameters (two hidden layers of 60 and 80 neurons, hyperbolic tangent activation, SGD optimizer, $L_2 = 24.45$) achieves the best performance overall ($4.24 \pm 0.19$\,kJ/mol), with a small gap between training and test errors confirming the absence of overfitting.

\begin{table}[ht]
    \centering
    \begin{tabular}{lll}\hline
    \textbf{Model}  & \textbf{Training (kJ/mol)} & \textbf{Test (kJ/mol)} \\\hline
    Linear          & $7.98 \pm 0.03$            & $6.09 \pm 0.21$        \\
    Random Forest   & $2.52 \pm 0.04$            & $4.86 \pm 0.29$        \\
    MLP             & $4.02 \pm 0.08$            & $4.24 \pm 0.19$        \\\hline
    \end{tabular}
    \caption{RMSE (kJ/mol) on training and test datasets for each supervised
    ML model, using the full descriptor set and 4-fold cross-validation on
    quinary systems.}
    \label{tab:RMSE}
\end{table}

\begin{figure}[ht]
    \centering
    \includegraphics[width=0.85\textwidth]{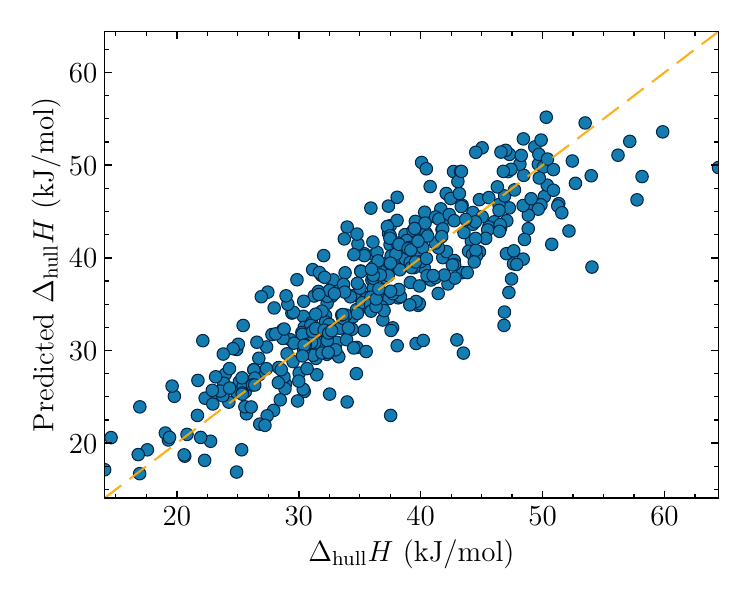}
    \caption{Predicted vs.\ DFT-calculated $\Delta_\mathrm{hull}H$ for the 416 quinary NaCl SQS mixtures. Each point corresponds to a quinary composition evaluated on its respective test fold. The dashed line represents a ideal prediction.}
    \label{fig:prediction-DhullH}
\end{figure}

The MLP is therefore selected for the large-scale screening thanks to the RMSE $\sim 4\,\text{kJ/mol}$.
Its predictions show suitable agreement with DFT reference $\Delta_\mathrm{hull}H$ values (Fig.~\ref{fig:prediction-DhullH}), with most quinaries scattering tightly around the ideal predicted versus DFT line.
A first comment about this result is that the prediction error is larger than typical DFT prediction based on ordered compounds as intermetallics where the RMSE could be below 2\,kJ/mol in some supervised ML trained on some recent curated database~\cite{zhang_prediction_2025}.
This may indicate that the nature of HEO is too complex for being estimated by simple descriptors since the nature of the oxygen bond with elements as transition metals is difficult to be expressed with simple GGA without Hubbard correction, in addition and mainly the main point of discussion here is that the mixing of cation elements leads to a disordered structure where the relaxation in DFT theme is not negligible and system dependent that could not be handled with our approximation in the ML models we use.
Nevertheless, RMSE of 4\,kJ/mol is still accurate enough to estimate a temperature within 600\,K and will be used for the following estimation of $T_\mathrm{stab}$ in the entire set of configurations. 

\begin{figure}[ht]
    \centering
    \includegraphics[width=0.85\textwidth]{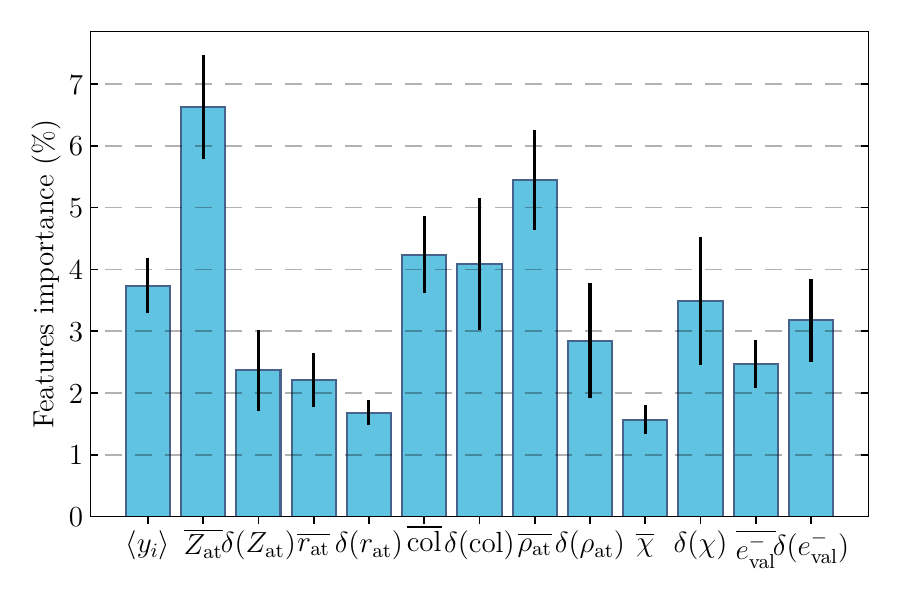}
    \caption{Relative importance of descriptors in the MLP model (mean $\pm$ standard deviation over the 4-fold cross-validation).}
    \label{fig:features-importance}
\end{figure}

The relative importance of descriptors in the MLP model (Fig.~\ref{fig:features-importance}) shows that the mean atomic number carries the highest weight, while most other features contribute comparably. 
This contrasts with the random forest, where mean atomic density dominates, reflecting different non-linear sensitivity patterns between the two model families.

\subsection{Most promising quinary ESO candidates}
\label{subsec:candidates}

\begin{table}[ht]
    \centering
    \begin{tabular}{ll}\hline
        \textbf{Compound} & $T_\mathrm{stab}$ (K) \\\hline
        (Ca,Cu,Fe,Mg,Zn)O  & 1,676 \\
        (Ca,Cu,In,Sr,Zn)O  & 1,750 \\
        (Ca,Cu,Ni,Sr,Zn)O  & 1,868 \\
        (In,Sn,Sr,V,Zn)O   & 2,080 \\
        (Ca,Cu,Fe,Ni,Zn)O  & 2,151 \\
        (Ca,Cu,In,Mg,Sr)O  & 2,153 \\
        (Ca,Co,Cu,Fe,Zn)O  & 2,183 \\
        (Ca,Co,Cu,Mg,Sr)O  & 2,214 \\
        (Ca,Co,Cu,Sr,Zn)O  & 2,281 \\
        (Ca,In,Mg,Sr,Zn)O  & 2,302 \\\hline
    \end{tabular}
    \caption{Ten quinary NaCl-type ESO candidates with the lowest predicted $T_\mathrm{stab}$ (GGA). 
    The full candidate list ($T_\mathrm{stab} < 3{,}500$\,K, 72 compounds) is given in the Supplementary Material~F.}
    \label{tab:promising-compound}
\end{table}

The MLP model is then applied to predict to all 4,368 possible equimolar quinary NaCl systems using the full training database.
All 10 quinary compositions with GGA-predicted $T_\mathrm{stab} < 3{,}500$\,K are listed in Table~\ref{tab:promising-compound} 
(full table in Supplementary Material~G). 

Despite the approximations inherent to the workflow (SQS disorder description, ideal $\Delta S_\mathrm{conf}$, neglect of magnetic and vibrational entropy, GGA-level large-scale screening), the model provides a raisonable relative ranking of synthesis targets.
Almost all candidates respect cation charge balance consistent with an average oxidation state of $+$2 compatible with the rocksalt structure (O at $-$2), demonstrating that the model correctly captures chemical constraints. 
Several compositions closely resemble the benchmark systems of Table~\ref{tab:promising-compound}, further reinforcing physical plausibility.

\newpage
\subsection{Experimental synthesis and Decomposition paths}
\label{subsec:decomp}



Building on the list of promising ESO candidates identified in the previous sections, we now confront the theoretical predictions with experimental synthesis tests and thermodynamic decomposition paths. 
Experimental work carried out by collaborators at ICMMO confirms partial NaCl-type multicomponent phase formation for several compositions, but also reveals cases where the samples melt at temperatures below the predicted stabilization temperature $T_\mathrm{stab}$, thereby preventing solid-state ESO formation. 
These observations highlight the need to account for liquid-phase stability in future iterations of the model and motivate a joint analysis of experimental data and computed decomposition paths.

In this subsection, we present three representative case studies for which experimental outcomes are interpreted in light of the computed finite-temperature convex-hull decomposition paths. 
For each composition, the convex-hull algorithm tracks the temperature evolution of all competing phase equilibria, enabling a direct comparison between predicted decomposition sequences and experimental observations. 
Unless otherwise stated, all thermodynamic quantities reported here are based on meta-GGA r$^2$SCAN calculations for the competing phases identified by the GGA-based screening.

\subsubsection{Experimental methodology}
All samples were synthesized by solid state reaction. 
Powder oxide precursors were weighted in stoichiometric amount, and mixed homogeneously using ball milling at 350\,rpm during 30\,minutes (Fritsch Pulverisette 7 premium line, with agate balls and vial). 
The resulting mixtures were pressed into pellets and annealed in alumina crucibles under air during 24\,h followed by quenching in liquid nitrogen. 
First synthesis attempts were performed at 1,173\,K, and then with increasing annealing temperatures until reaching the melting point. 
X-ray diffraction patterns were recording using a PANalytical X’Pert diffractometer equipped with a Pixel detector. 
Chemical mapping were obtained using a scanning electron microscope (SEM) with energy-dispersive spectroscopy (EDS), using an SEM-FEG Zeiss Sigma HD. 
Melting points were measured by DSC using a SETARAM LabSys instrument.

\subsubsection{(Cu,Fe,Mg,Sr,Zn)O} 
\begin{figure}
          \centering
       \includegraphics[width=0.5\textwidth]{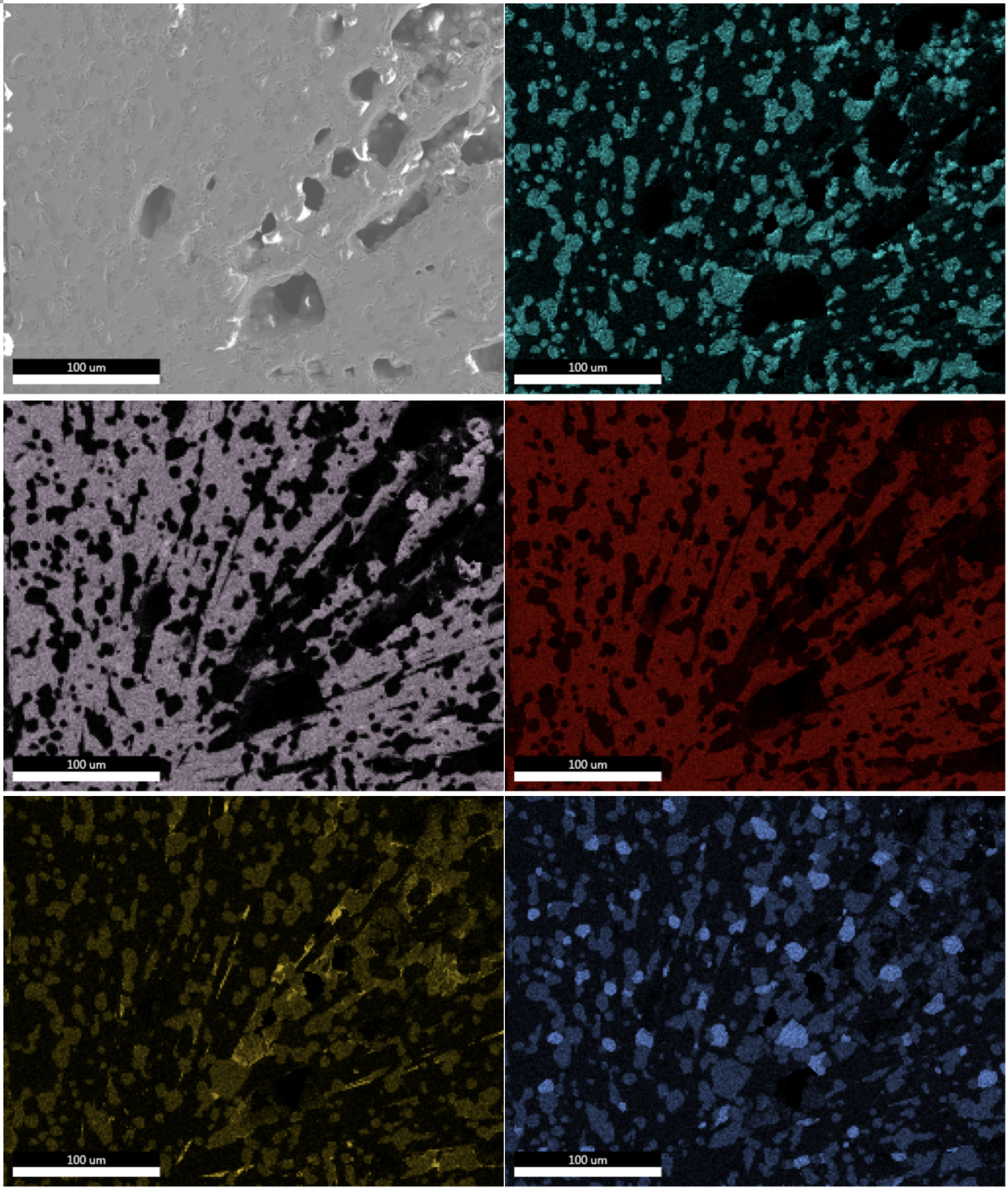}
        \caption{Top left: SEM picture of Cu–Fe–Mg–Sr–Zn–O sytem, top right: Mg mapping, middle left: Sr mapping, middle right: Fe mapping, bottom left: Cu mapping, bottom right: Zn mapping.}
          \label{fig:exp_case1}
     \end{figure}

This composition starts melting incongruently at 1,333\,K.
At 1,173\,K, the sample consists of a mixture of several phases, mostly a rocksalt phase with lattice parameters close to MgO, a strontium ferrite SrFeO$_{3-x}$, Cu$_3$MgO$_4$, and minor amount of ZnO and unidentified phases. 
At larger temperatures, part of these phases disappear, but the compounds remains multiphase, mostly consisting of a MgO-like rocksalt phase, SrFeO$_{3-x}$ and ZnO. 
These results are confirmed by chemical mapping performed by by SEM-EDX (Fig.~\ref{fig:exp_case1}). They show an obvious segregation of the cations, with Sr and Fe forming a binary oxide with not solubility of Mg and Cu but partial solubility of Zn.

\begin{figure}
          \centering
        \includegraphics[width=0.8\textwidth]{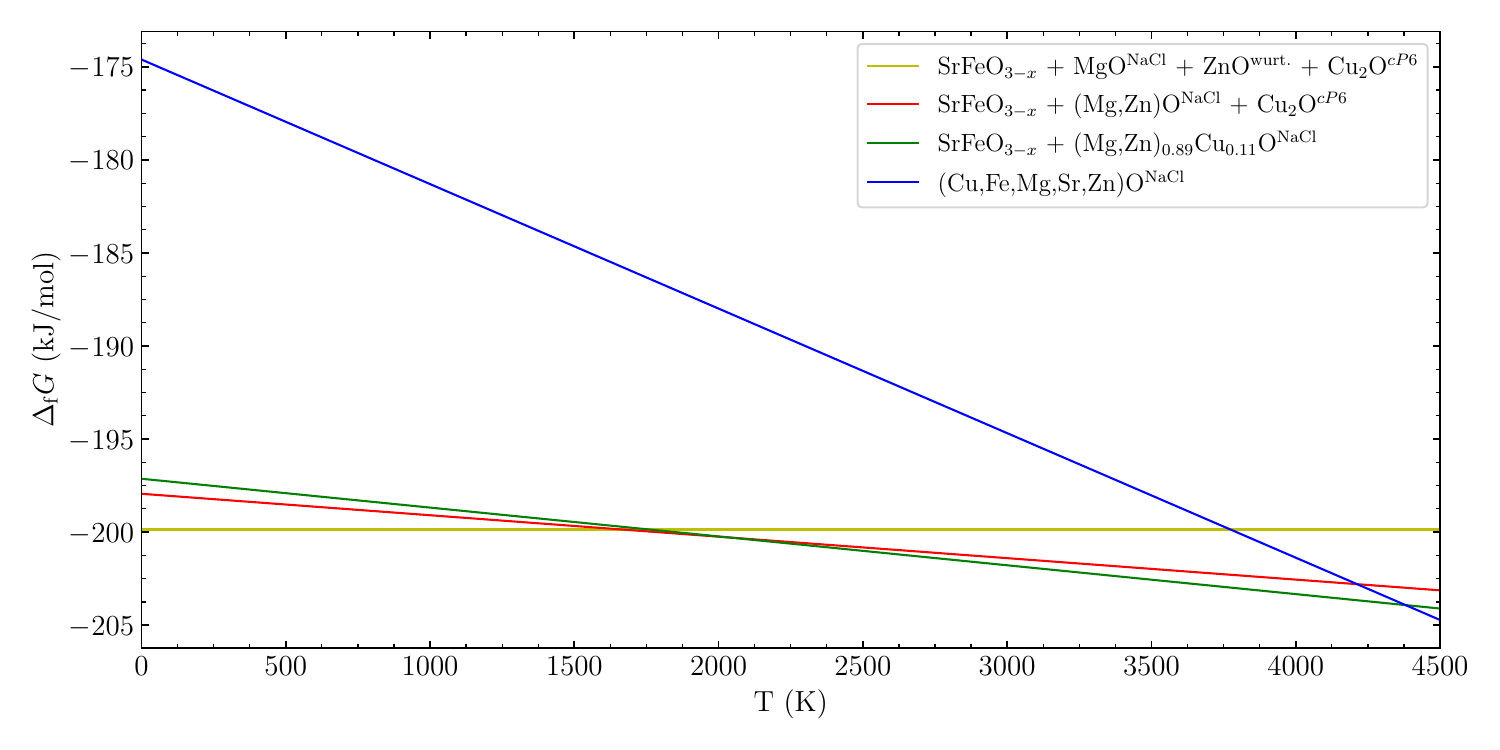}
        \caption{Gibbs energy compute with meta-GGA or some phases of the Cu–Fe–Mg–Sr–Zn–O system as a function of temperature using meta-GGA.}
          \label{fig:stab_SrFe}
     \end{figure}

About the computational approach, the ternary SrFeO$_{3-x}$-type perovskite phase was initially omitted from the reference database; once computed and added, it proved highly stable, causing a significant upward revision of $T_\mathrm{stab}$ for all compositions containing both Sr and Fe.
The predicted phase equilibria are shown in Figure~\ref{fig:stab_SrFe} and summarized in Table~\ref{tab:stab_SrFe}. 

At low temperature, the most stable decomposition involves Sr$_4$Fe$_4$O$_{11}$, MgO, ZnO, and Cu$_2$O. 
A (Mg,Zn)O binary NaCl mixture appears above $\sim$1,663\,K, followed by a ternary (Cu,Mg,Zn)O NaCl mixture above 2,036\,K. 
The full quinary ESO only becomes thermodynamically stable at 4,381\,K, which is unachievable in practice. 
This analysis therefore confirms the experimental observation:
the NaCl phase that forms excludes Fe and Sr, and is consistent with the model predictions for the reduced quaternary subsystem, which stabilizes near 1,633\,K.
This case also highlights a methodological lesson: the presence of very stable ternary phases such as SrFeO$_{3-x}$ must be included in the reference database, otherwise $T_\mathrm{stab}$ will be systematically underestimated for all related compositions.

\begin{table}[ht]
\centering
\renewcommand{\arraystretch}{1.2}

\begin{tabular}{rl}\hline
$\bm{T}$ \textbf{(K)} & \textbf{Decomposition} \\\hline
0       & 0.475\,Sr$_4$Fe$_4$O$_{11}$ + 0.2\,MgO$^\mathrm{NaCl}$ +
          0.2\,ZnO$^\mathrm{wurt}$ + 0.075\,Cu$_2$O + 0.05\,Cu \\
1,663   & 0.475\,Sr$_4$Fe$_4$O$_{11}$ + 0.4\,(Mg,Zn)O$^\mathrm{NaCl}$ +
          0.075\,Cu$_2$O + 0.05\,Cu \\
2,036   & 0.475\,Sr$_4$Fe$_4$O$_{11}$ + 0.3\,(Mg,Zn)O$^\mathrm{NaCl}$ +
          0.15\,(Cu,Mg,Zn)O$^\mathrm{NaCl}$ + 0.075\,Cu \\
4,381   & (Cu,Fe,Mg,Sr,Zn)O$^\mathrm{NaCl}$ \\\hline
\end{tabular}
\caption{Temperature-dependent phase decomposition for the
Cu--Fe--Mg--Sr--Zn--O system (meta-GGA).}
\label{tab:stab_SrFe}
\end{table}

\subsubsection{(Ca,Co,Cu,Mg,Zn)O.} 
\begin{figure}
        \centering
        \includegraphics[width=0.7\textwidth]{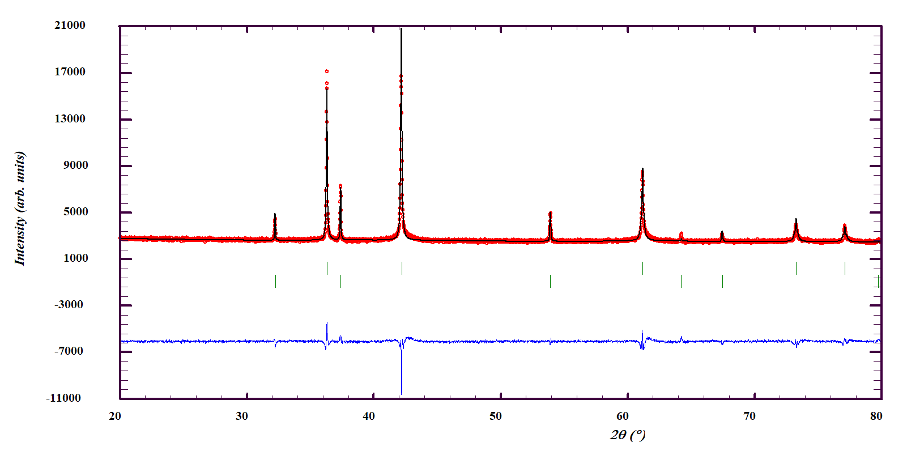}
        \caption{XRD pattern of (Ca,Co,Cu,Mg,Zn)O annealed at 1,223\,K, evidencing the presence of two rocksalt phases (note that the quality of the refinement is poor due to the Jahn-Teller distortion usually observed in high-entropy or medium-entropy rocksalt oxides containing Cu)}
       \label{fig:exp_case2a}
    \end{figure}

This composition starts melting incongruently at 1,453\,K. 
At 1,173\,K, the sample consists of a mixture of several phases, two rocksalt phases, Ca$_2$CuO$_3$, as well as unidentified phases. 
From 1,223\,K and up to the melting point, two phases are mostly observed, both with rocksalt crystal structure and very distinct lattice parameters, namely 4.280\,\AA\ and 4.804\,\AA\ (Fig.~\ref{fig:exp_case2a}). 
The largest value is rather close to the lattice parameter of calcium oxide (4.811\,\AA), whereas the smallest one is only slightly larger than the value obtained for the 4-cations compound (CoCuMgZn)O (4.253\,\AA) despite the very large radius of Ca$^{2+}$ as compared to the other cations. 
It shows that at least up to 1,453\,K, there is a limited solubility of Ca$^{2+}$ in (CoCuMgZn)O~\cite{mnasri_2021}. 

\begin{figure}
        \centering
        \includegraphics[width=0.8\textwidth]{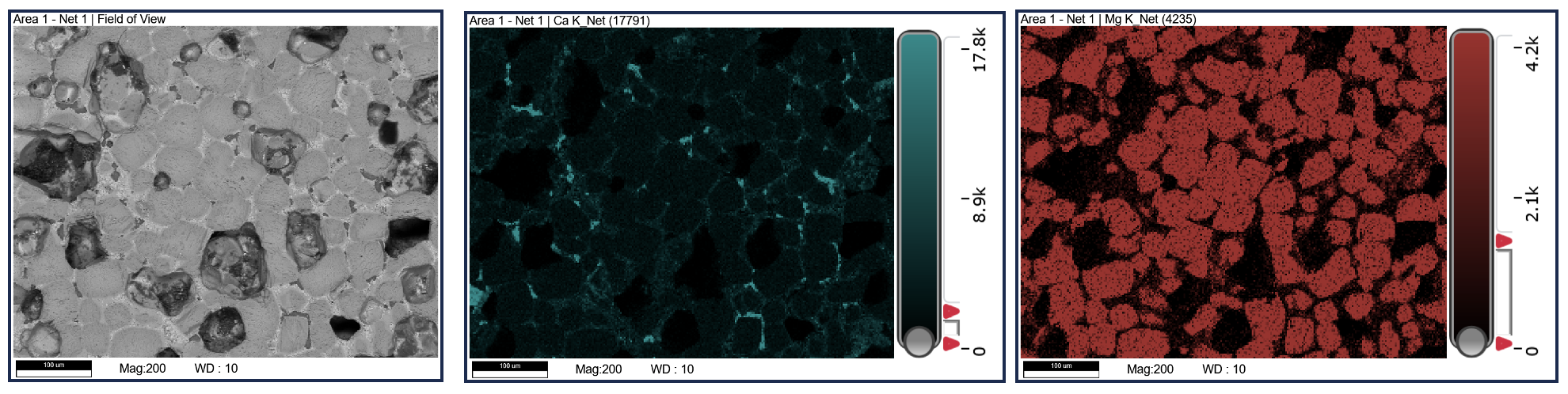}
        \caption{SEM picture of (Ca,Co,Cu,Mg,Zn)O system, middle: Ca mapping, right: Mg mapping.}
       \label{fig:exp_case2b}
    \end{figure}

\begin{figure}
        \centering
        \includegraphics[width=0.8\textwidth]{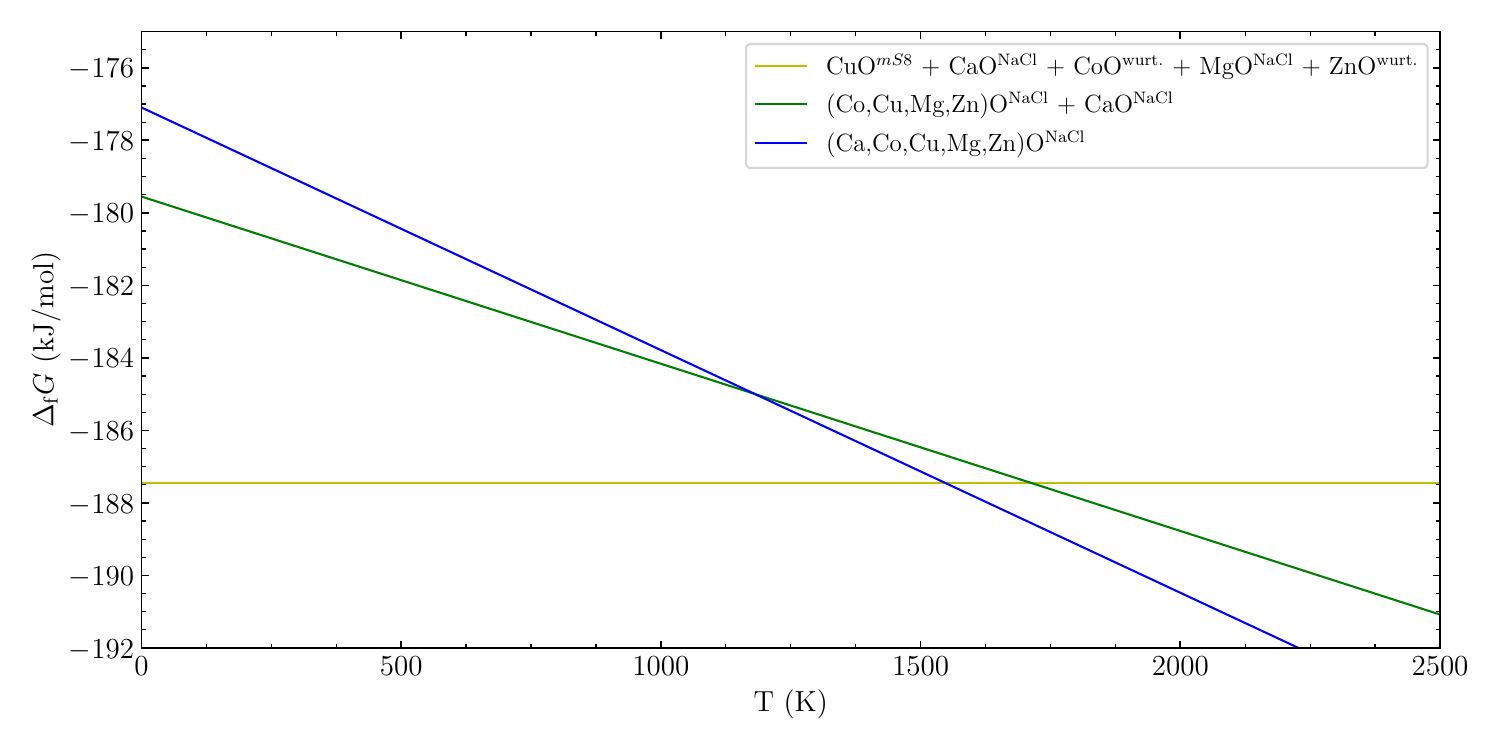}
        \caption{Gibbs energy compute with meta-GGA or some phases of the Ca–Co–Cu–Mg–Zn–O system as a function of temperature using meta-GGA.}
       \label{fig:stab_CaCo}
    \end{figure}

Thus, another synthesis has been attempted with a nominal composition Ca$_{0.1}$Co$_{0.225}$Cu$_{0.225}$Mg$_{0.225}$Zn$_{0.225}$. 
The results were similar, except a decrease of the proportion of the rocksalt phase with the largest lattice parameter, which confirms that it mostly consists of calcium oxide. 
To further validate this conclusions, chemical mappings were performed, evidencing an obvious segregation of calcium as compared to the other cations (see Figs~\ref{fig:exp_case2b} and \ref{fig:stab_CaCo}).

The GGA-predicted $T_\mathrm{stab}$ for the quinary ESO was 2,673\,K.
With meta-GGA and inclusion of quaternary competing phases, the quinary ESO stabilizes near 1,500\,K in the model. 
Below this temperature, it decomposes into five ordered binary oxides. 
The free energy of the phase equivalent to the experimentally observed ESO (without Ca) is found to be very close to, but never below, the quinary hull near its predicted $T_\mathrm{stab}$, consistent with the experimental finding that Ca segregates into a separate CaO-rich NaCl phase. 
The discrepancy between model and experiment is of order a few kJ/mol, which can in part be attributed to stoichiometry deviations not captured by the ideal-composition DFT model. 
Overall, the agreement is satisfactory.

\subsubsection{(Ca,Cu,Mg,Ni,Zn)O}

\begin{figure}
        \centering
        \includegraphics[width=0.7\textwidth]{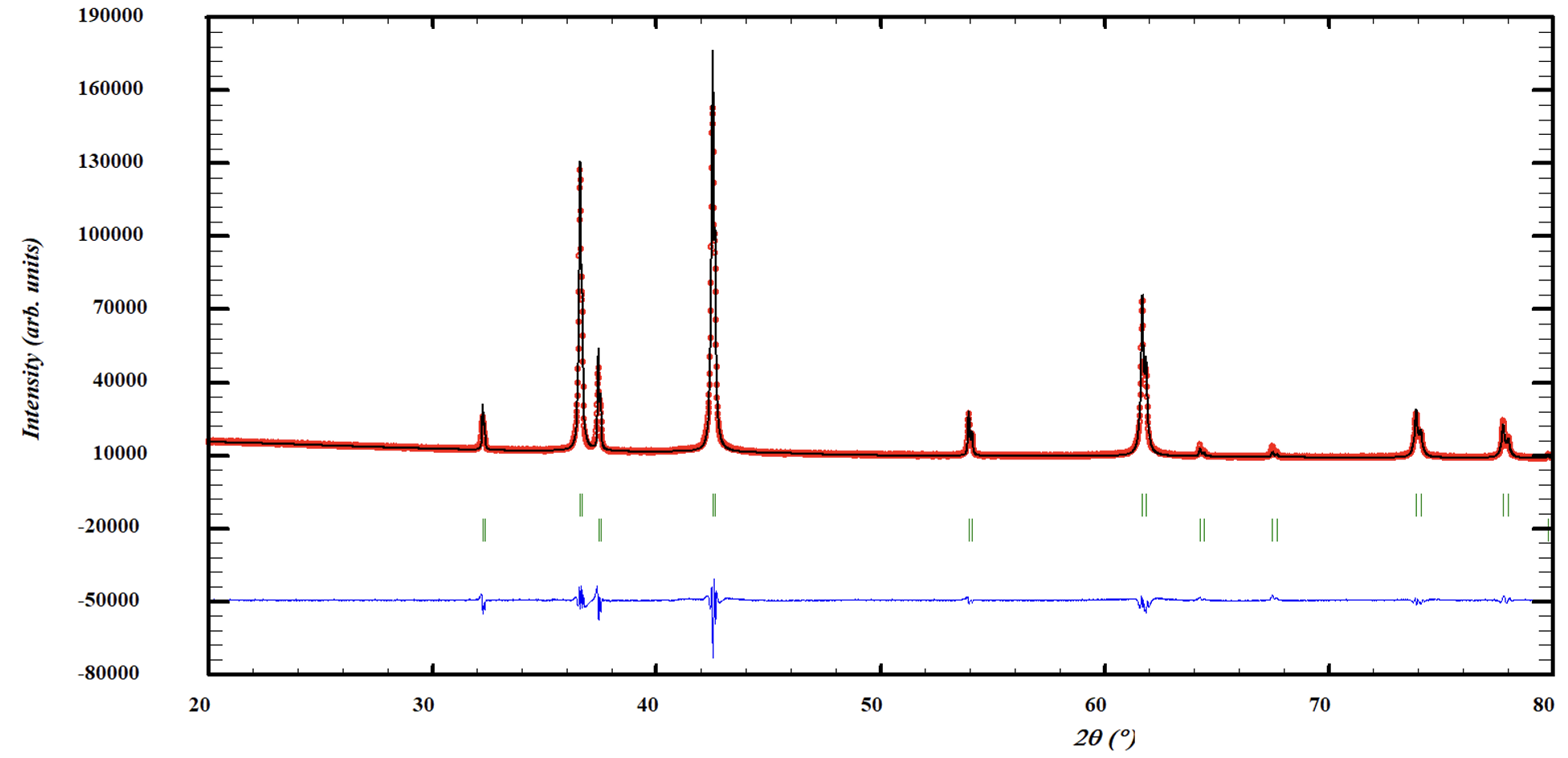}
        \caption{XRD pattern of (Ca,Cu,Mg,Ni,Zn)O annealed at 1,323\,K, evidencing the presence of two rocksalt phases (note that the quality of the refinement is poor due to the Jahn-Teller distortion usually observed in high-entropy or medium-entropy rocksalt oxides containing Cu)}
        \label{fig:exp_case3a}
    \end{figure}

The experimental observations are very similar to the case of (Ca,Co,Cu,Mg,Zn)O. 
This composition starts melting incongruently at between 1,423\,K and 1,473\,K. 
At 1,173\,K, the sample consists of a mixture of several phases, two rocksalt phases, calcium cuprate Ca$_{1-x}$CuO$_2$, ZnO, as well as unidentified phases. 
Part of these phases disappear at 1,223\,K, and from 1,323\,K up to the melting point only two rocksalt phase remain with very distinct lattice parameters, namely 4.249\,\AA\ and 4.803\,\AA\, as shown in Fig~\ref{fig:exp_case3a}. 

\begin{figure}
        \centering
        \includegraphics[width=0.8\textwidth]{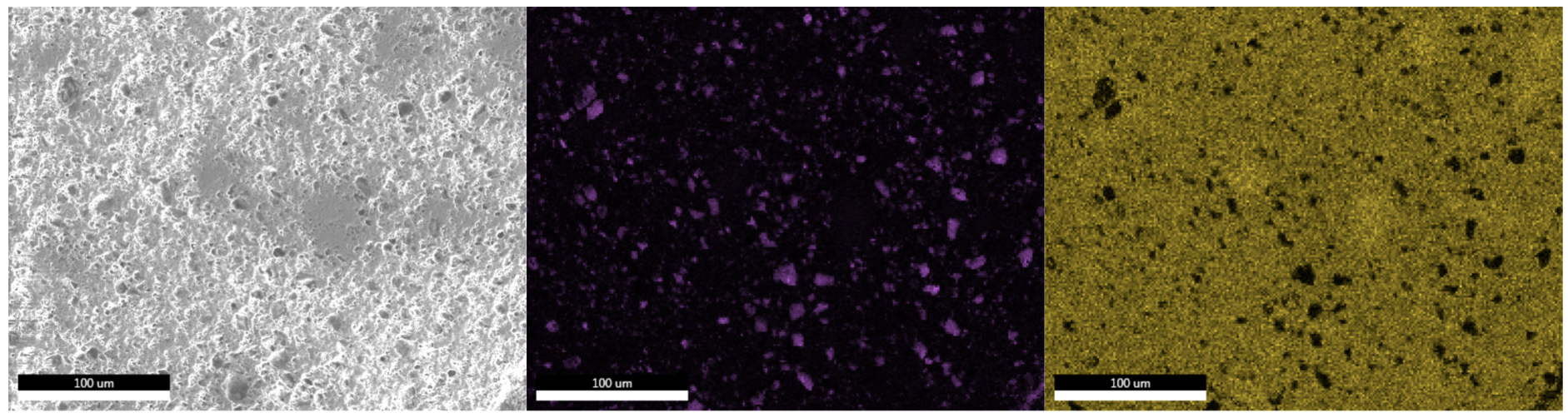}
        \caption{Left: SEM picture of (Ca,Cu,Mg,Ni,Zn)O, middle: Ca mapping, right: Cu mapping
}
        \label{fig:exp_case3b}
    \end{figure}

Here again, the largest value is rather close to the lattice parameter of calcium oxide, whereas the smallest one is only slightly larger than the value obtained for the 4-cations compound (CuMgNiZn)O (4.234\,\AA{})~\cite{mnasri_2021}, which evidences a limited solubility of Ca$^{2+}$ in this 4-cations compound. 
This conclusion was confirmed by chemical mapping evidencing a clear segregation of Ca (Fig.~\ref{fig:exp_case3b}). 
Similarly to the previous case, another synthesis has been attempted with a nominal composition Ca$_{0.1}$Cu$_{0.225}$Mg$_{0.225}$Ni$_{0.225}$Zn$_{0.225}$ leading to the same observation.

\begin{figure}
        \centering
        \includegraphics[width=0.8\textwidth]{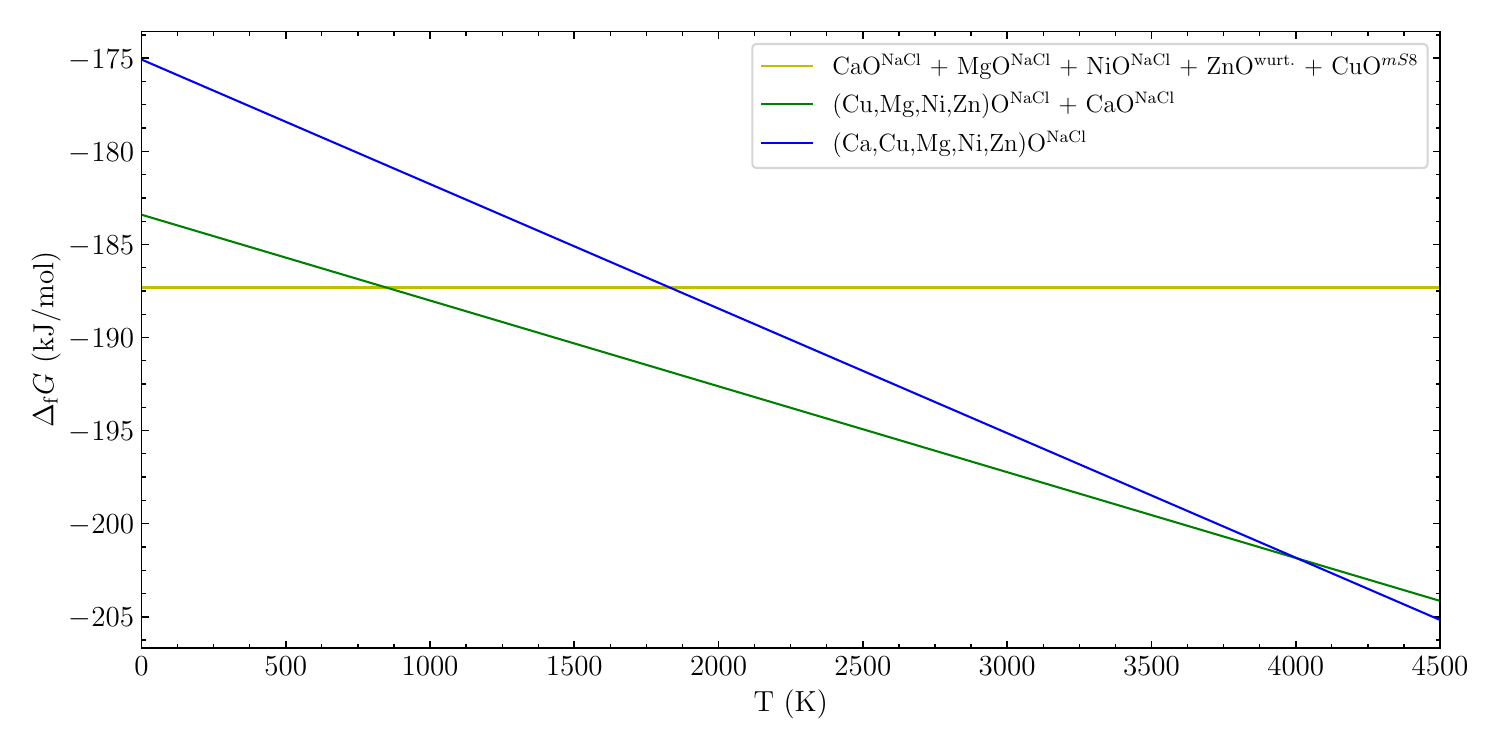}
        \caption{Gibbs energy compute with meta-GGA or some phases of the Ca–Cu–Mg–Ni–Zn–O system as a function of temperature using meta-GGA.}
        \label{fig:stab_CaCu}
        \end{figure}

About the comparison with the computational decomposition path, the
meta-GGA calculations predict formation of (Cu,Mg,Ni,Zn)O + CaO above $\sim$1,550\,K. 
The full equimolar quinary is only predicted to stabilize above 4,000\,K, consistent with the experimental observation that Ca does not incorporate into the ESO under accessible synthesis conditions (Fig.~\ref{fig:stab_CaCu}).

\subsubsection{Final remarks}

These three case studies demonstrate that our theoretical model correctly captures the tendency of Ca (and other elements with strongly differing size or chemistry) to segregate rather than fully homogenize into the quinary NaCl phase.
They also suggest that departing from strict equimolarity, for instance by slightly reducing the Ca fraction, may offer a practical route to obtain genuine quinary ESOs at accessible synthesis temperatures. 
The compositions predicted in the full candidate table (Supplementary Material) thus represent good starting points for non-equimolar ESO exploration, excluding those that melt prematurely.

\changed{We also tested other samples as two In‑containing quinary compositions, (Ca,Cu,In,Sr,Zn)O and (Ca,Cu,In,Mg,Sr)O.
Under the present solid‑state synthesis conditions (1223–1373 K), these did not yield single‑phase rocksalt structure but multiphase mixtures dominated by In‑rich phases (e.g. SrIn$_2$O$_4$ / CaIn$_2$O$_4$ type). 
This behavior is in line with our theoretical predictions, which indicate a low likelihood of stabilizing rocksalt structures with In$^3+$ under conventional synthesis routes. 
These results will be detailed in a future, more experimentally focused study.}


\section{Conclusion}

In this work, we developed a computational framework to screen equimolar quinary entropy-stabilized oxides (ESOs) in the NaCl structure type by combining SQS-based DFT calculations, convex-hull thermodynamics, and supervised machine learning. 
The workflow provides a reliable relative ranking of candidate stability: known NaCl-type ESOs are recovered among the lowest-$T_\mathrm{stab}$ compositions, and the predicted decomposition products are consistent with the experimentally observed secondary phases. 
This supports the use of the model as an effective screening tool for prioritizing promising compositions and rationalizing experimental outcomes.

From a methodological perspective, the optimized neural-network model predicts the thermodynamic properties of the 4,368 quinary compositions with high accuracy while requiring DFT calculations for only about 10\% of them. 
By contrast, simpler linear estimates of $\Delta_\mathrm{mix}H$ reproduce only broad trends and remain insufficient for predictive screening. 
These results highlight the importance of nonlinear correlations in governing the stability of multicomponent oxides.

At the same time, absolute values of $T_\mathrm{stab}$ remain affected by systematic uncertainties related to the use of GGA in the large-scale screening, the neglect of magnetic entropy, the assumption of ideal stoichiometry, the incompleteness of the ordered reference database, and the absence of liquid-phase thermodynamics. 
They should therefore be interpreted as comparative indicators rather than exact synthesis temperatures, although r$^2$SCAN significantly improves the energetic description.

More broadly, the results suggest that fully single-phase equimolar quinary NaCl-type ESOs remain difficult to obtain and that their stability is not governed by configurational entropy alone. 
Instead, enthalpic effects such as charge transfer, oxidation-state compatibility, and competition with lower-order oxides appear to play a central role. 
Overall, this work provides an efficient route for the data-driven exploration of multicomponent oxides and a practical framework for guiding future experimental investigations.

\section*{Acknowledgments}
This research is part of the Detox Project (ANR-21-CE08-0006-01). 
DFT calculations were performed using HPC resources from GENCI–CINES (Grant A0060906175). 
We acknowledge Center for Computational Materials Science, Institute for Materials Research, Tohoku University for the use of MASAMUNE-II (Project No.2412SC0510)

\section*{Code Availability}
The Iterative convex-hull implementation based on the Quickhull algorithm is available at \url{https://github.com/link-lab3629}
. Scripts used for postprocessing will be deposited together with the data (calculated training database and prediction of the whole combination of HEO) in the github after publication.

\section*{Supplementary Materials}
\begin{itemize}
    \item[A] SQS and DFT parameters 
    \item[B] Ground-state of binary $M_1 -$O systems 
    \item[C] Details of ternary $M_1 - M_2 -$O ordered compounds
    \item[D] Training database analysis 
    \item[E] Performance of machine learning models
    \item[F] Prediction of the whole combination of HEO
\end{itemize}

\bibliographystyle{elsarticle-num-names} 
\bibliography{bibliographie}



\section*{Supplementary material}

# Supplementary Materials of the Paper: Data-Driven Prediction of NaCl-Type Entropy-Stabilized Oxide Compositions from First-Principles and Supervised Learning.

Sébastien Junier[a], Céline Barreteau[a], David Beradan[c], Yann-Andrev Kerneur[c], Jean-Claude Crivello[b]

[a] *Univ Paris Est Creteil, CNRS, ICMPE, UMR 7182, 2 rue Henri Dunant, Thiais, 94320, France*
[b] *CNRS – Saint-Gobain – NIMS, IRL3629, Laboratory for Innovative Key Materials and Structures (LINK), National Institute for Materials Science, 1-1 Namiki, Tsukuba, Ibaraki, 305-0044, Japan*
[c] *Université Paris-Saclay, CNRS, Institut de chimie moléculaire et des matériaux d'Orsay, Orsay, 91405, France*

## Appendix A. SQS and DFT parameters

The RMSE of cluster correlation function and DFT total energy as a function of the number of atoms in the SQS supercell and SQS-supercell for binary, ternary, and quinary HEO mixture are given in Fig. A.1.

The following structures have been kept. The SQS POSCAR files could be download from the following links:

- 32 atoms for binary $(M_1, M_2)$O mixtures (8 cations per species + 16 O), with a residual energy uncertainty of $\sim$2 kJ/mol;
  `https://sdrive.cnrs.fr/s/ZMGQEMaBG93FWTq`
- 48 atoms for ternary $(M_1, M_2, M_3)$O mixtures (8 cations per species + 24 O), with $\sim$1 kJ/mol uncertainty;
  `https://sdrive.cnrs.fr/s/76CPJMCFRtcYfMa`
- 60 atoms for equimolar quinary $(M_1, \ldots, M_5)$O mixtures (6 cations per species + 30 O), with $\sim$1 kJ/mol uncertainty.
  `https://sdrive.cnrs.fr/s/acCS2sdHyg2kHQC`

For each structures, the uncertainty due to $\vec{k}$-point meshing is below $10^{-2}$ kJ/mol, as shown in Fig. A.2.

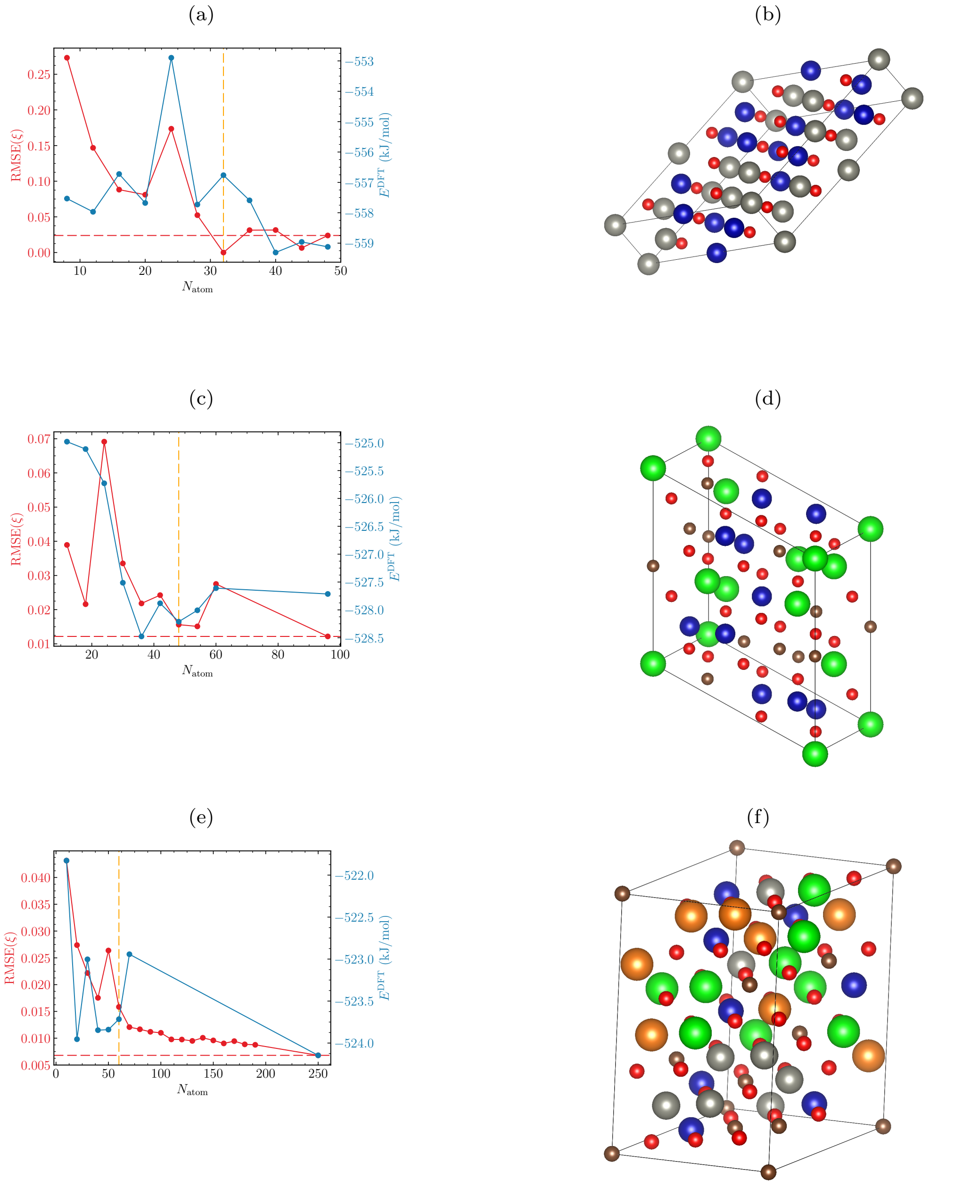


Figure A.1: RMSE of cluster correlation function and DFT total energy as a function of the number of atoms in the SQS supercell and SQS-supercell of (a,b) a binary, (c,d) ternary, and (e,f) quinary mixture for $(\{M_i\})O$ mixture of $M_i$ elements in NaCl structure.

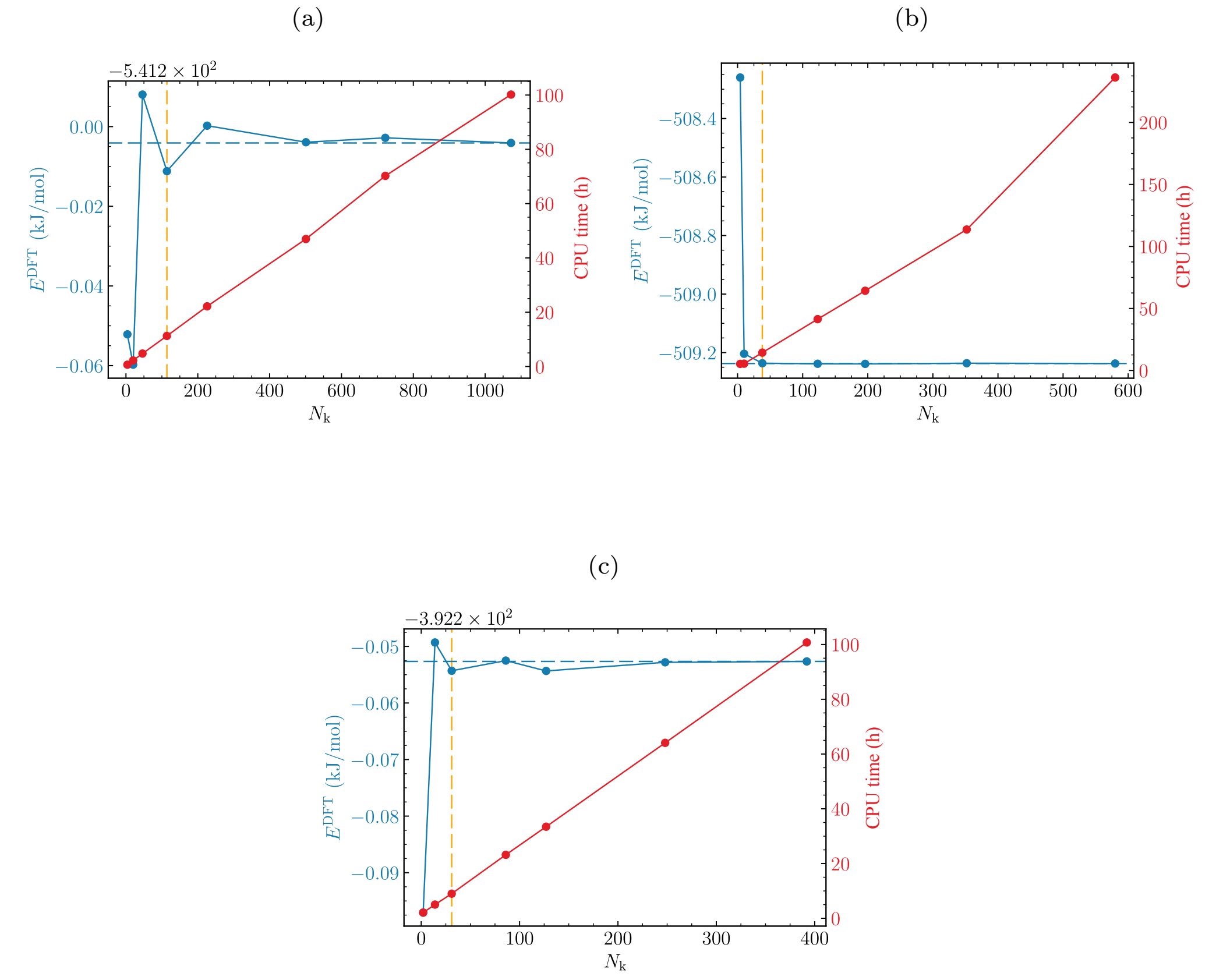


Figure A.2: DFT total energy and CPU computation time as functions of the number of $\vec{k}$-points for (a) binary, (b) ternary, and (c) quinary SQS supercells in the NaCl structure.

## Appendix B. Ground-state of binary $M_1$−O systems

For each $M_1$−O system, the formation enthalpy ($\Delta_\mathrm{f} H$) of all ordered oxides is plotted as a function of the oxygen concentration. A convex hull is then constructed to determine the ground state of the system at 0 K. The phases lying on this hull are therefore defined as stable at 0 K. Over the portion of the curve not defined by a compound, equilibrium exists between the two closest phases on either side. To facilitate comparison between the

different functionals, the formation enthalpy of each compound is shown on the same graph, both in GGA (filled circle, ●) and in meta-GGA (triangle, ▲). The GGA convex hull is shown as a dashed line, and the meta-GGA hull as a solid line. The phases expected to be stable according to experimental phase diagrams are circled in black (●). By definition, they should lie as close as possible to the convex hull. For all these values, the ZPE correction is applied. Calculations without ZPE correction are shown only through the GGA and meta-GGA convex-hull curves, displayed in red using the same line style as for the corrected hulls. In the following plots, experimental formation-enthalpy data are added in blue for comparison. These data were collected from the following references: `Cha98` [1], `Cox84` [2], `Bak71` [3], `Hun00` [4], `Kor75` [5], `Hol52` [6], `Rob95` [7], and `Mor93` [8].

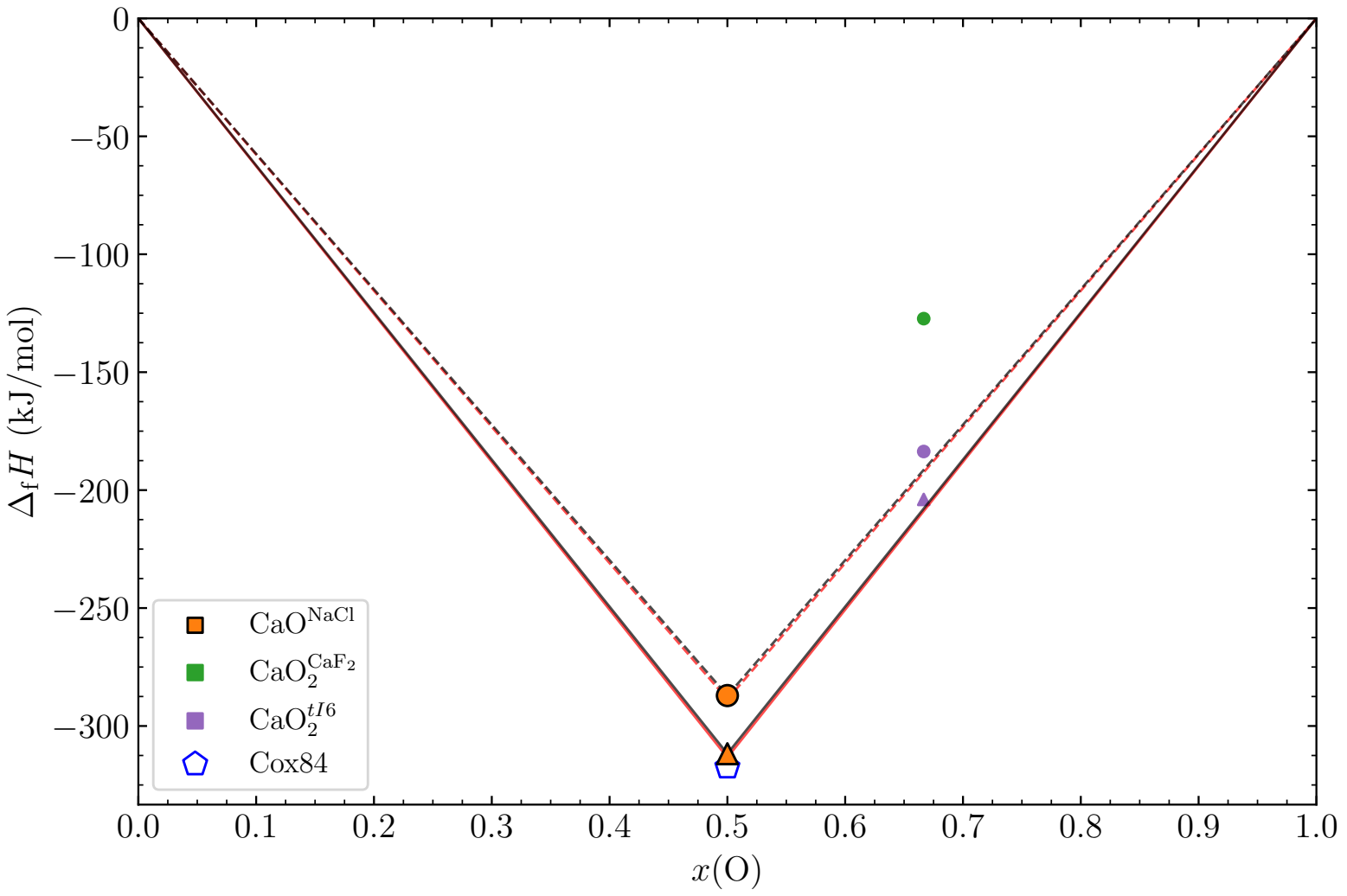


Figure B.1: Ground state of Ca−O system

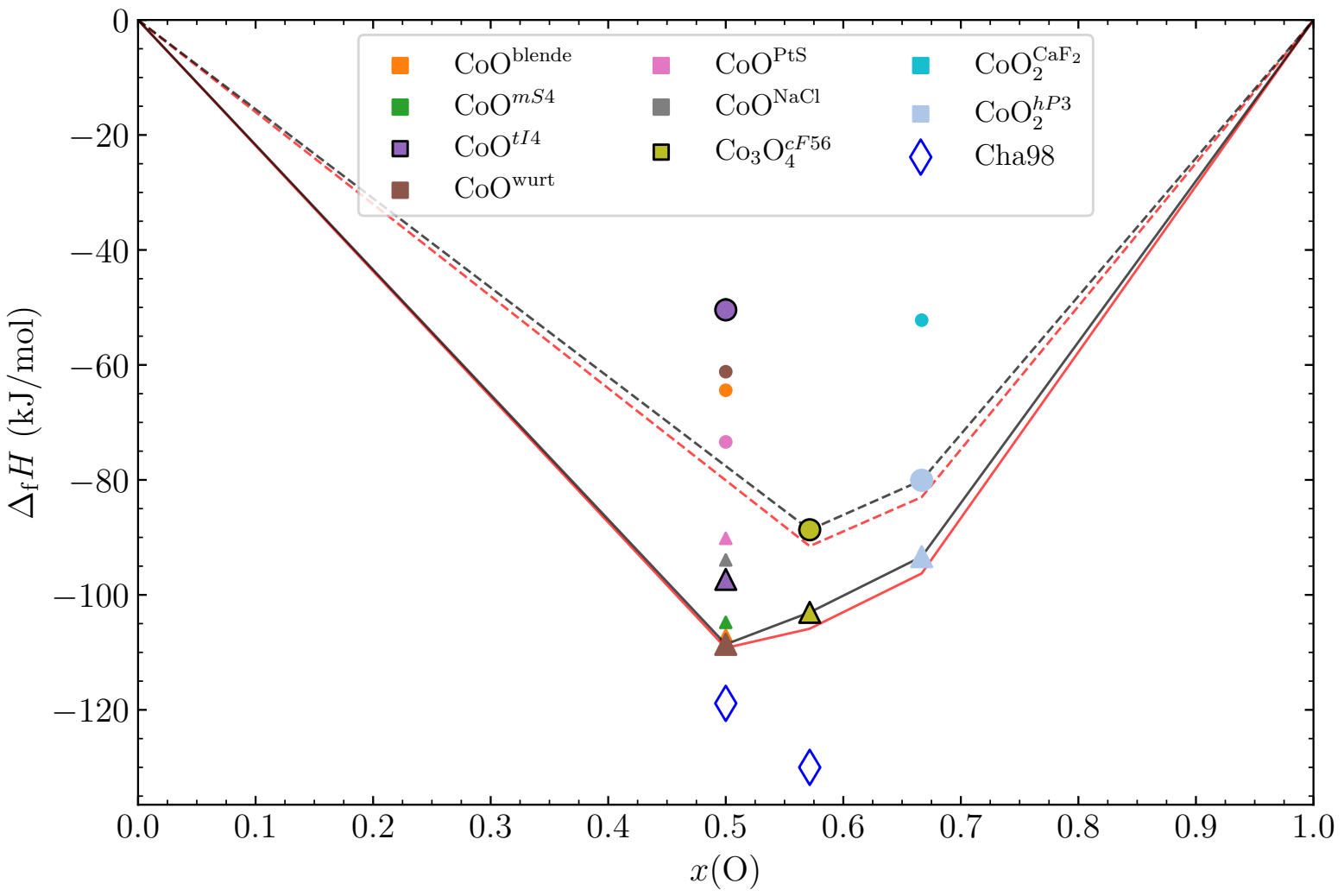


Figure B.2: Ground state of Co−O system

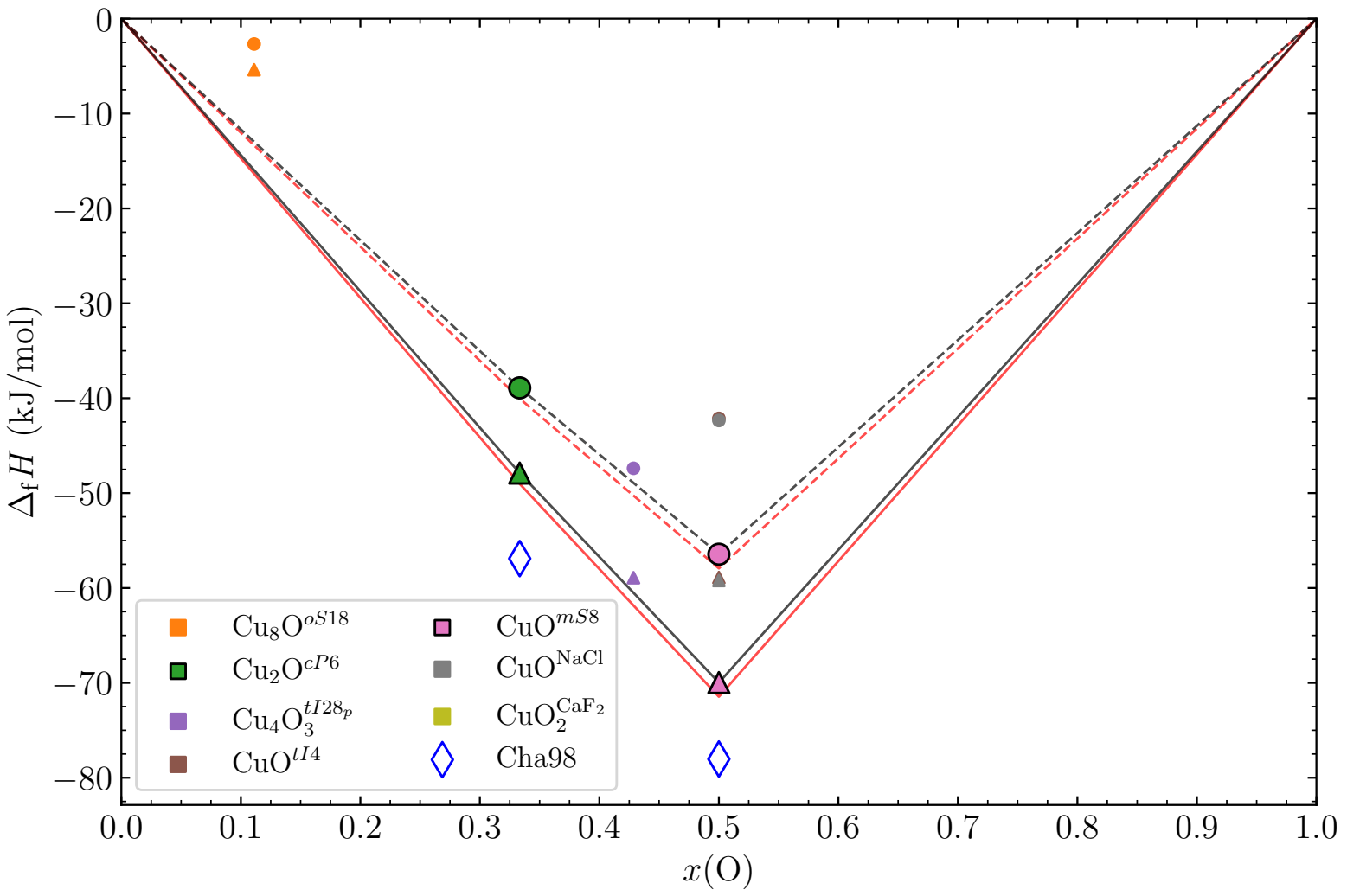


Figure B.3: Ground state of Cu−O system

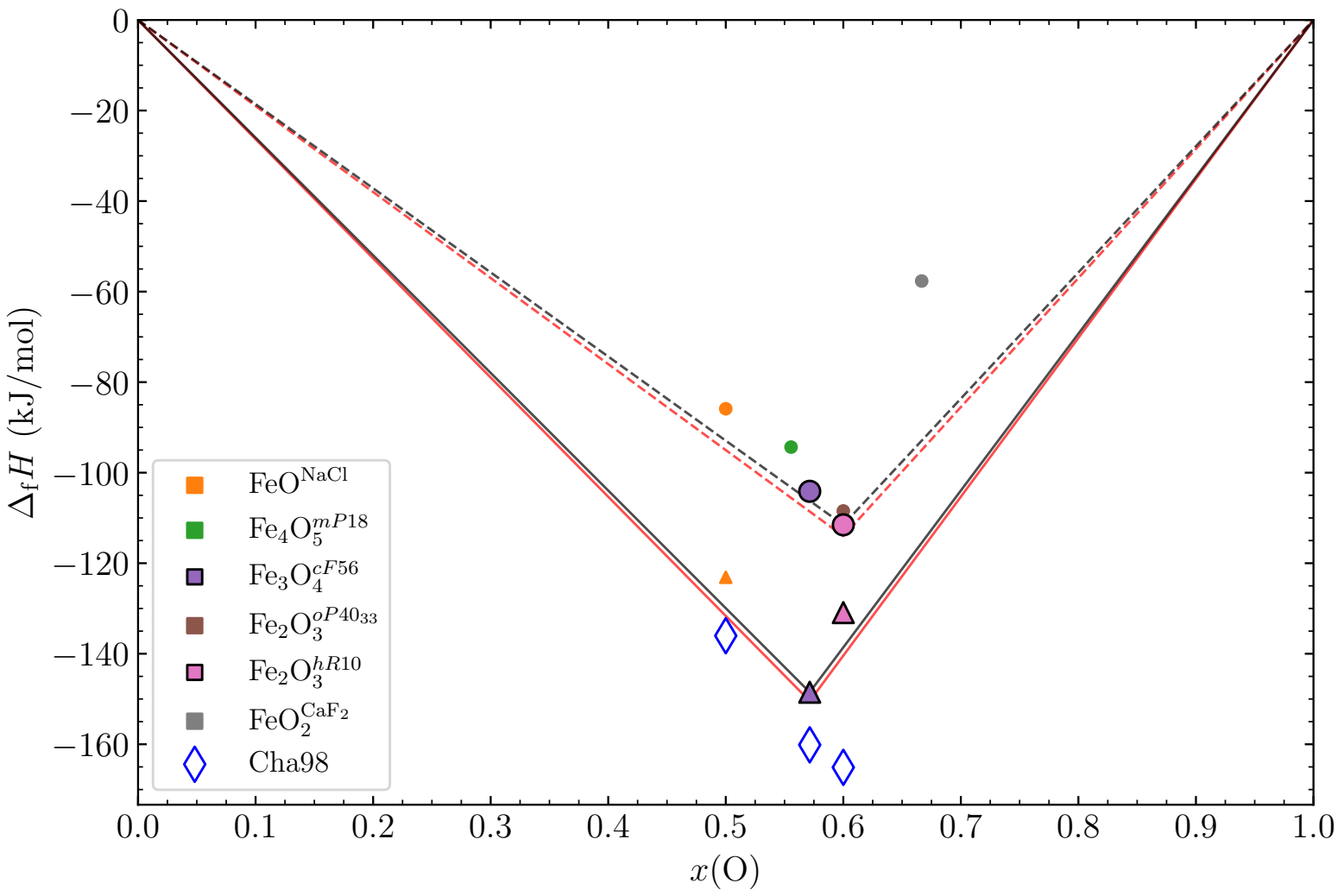


Figure B.4: Ground state of Fe−O system

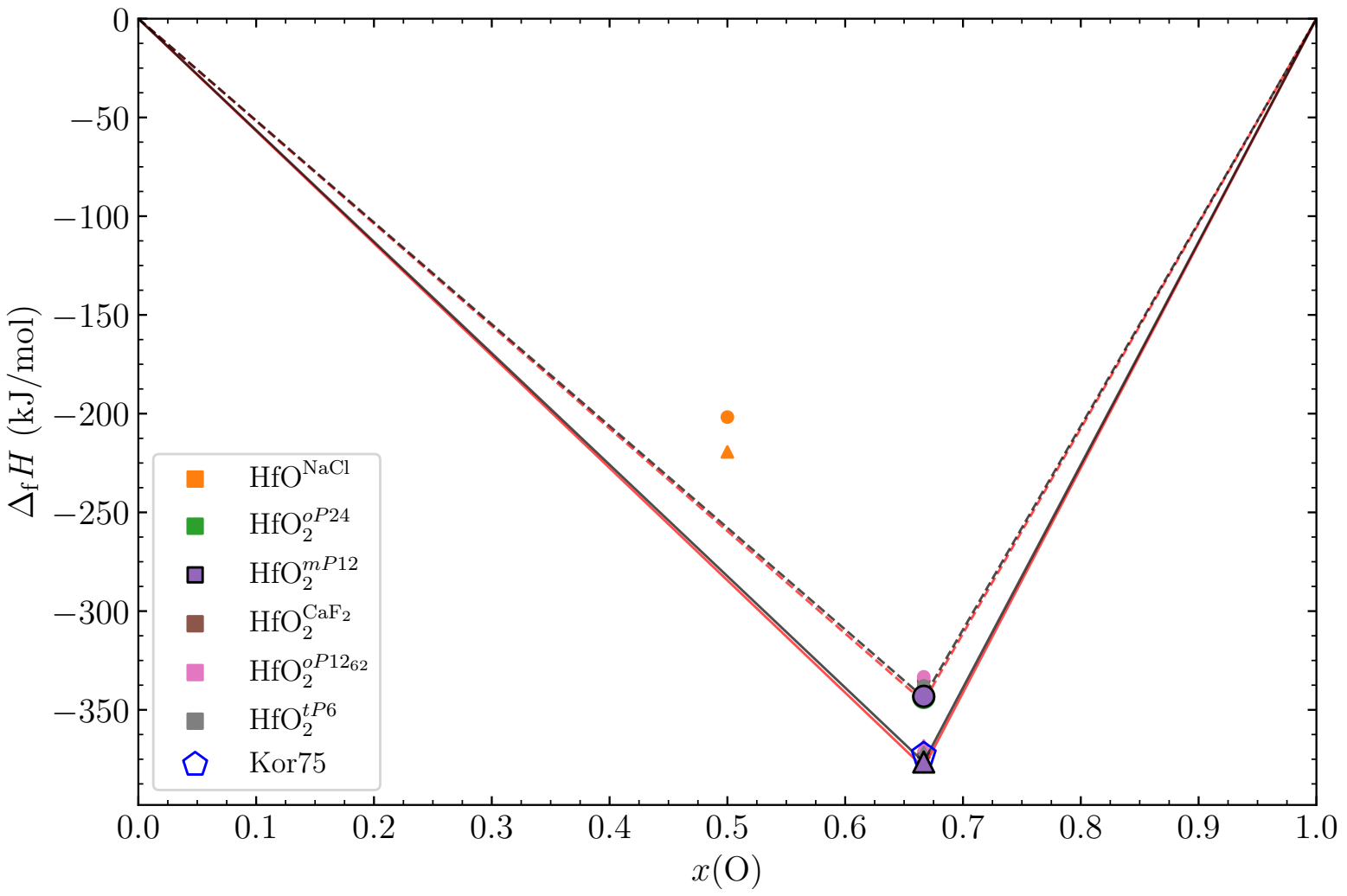


Figure B.5: Ground state of Hf−O system

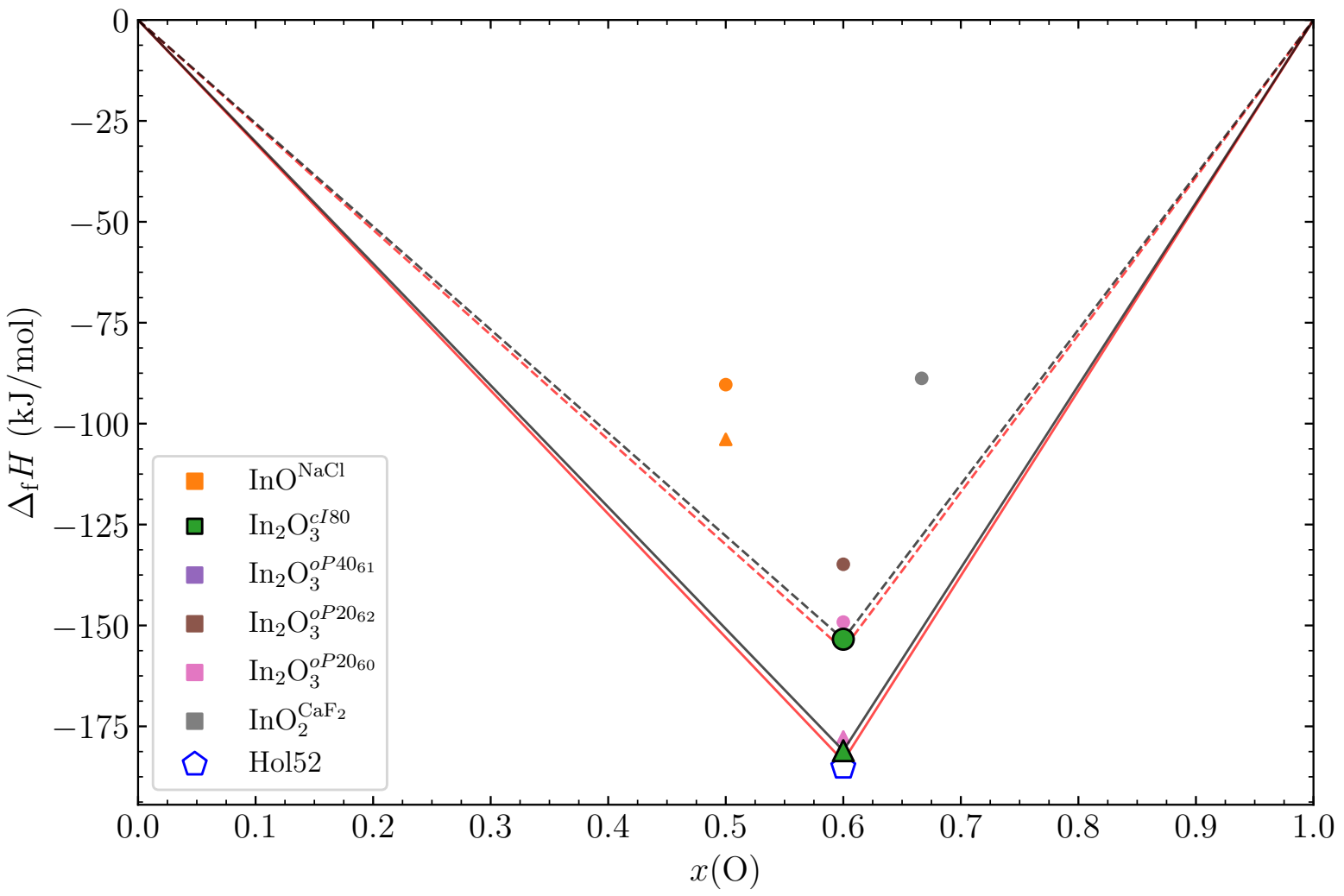


Figure B.6: Ground state of In−O system

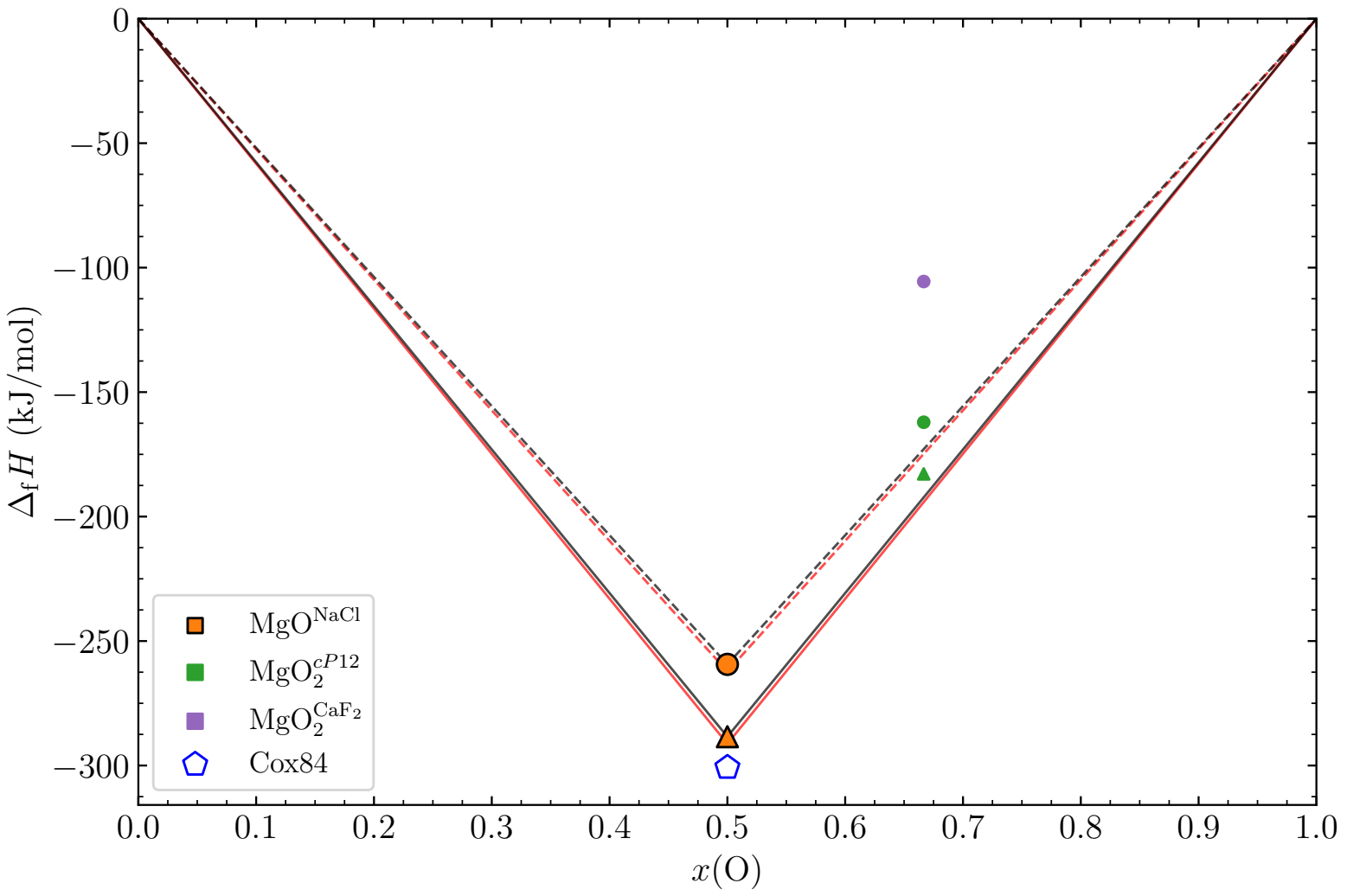


Figure B.7: Ground state of Mg−O system

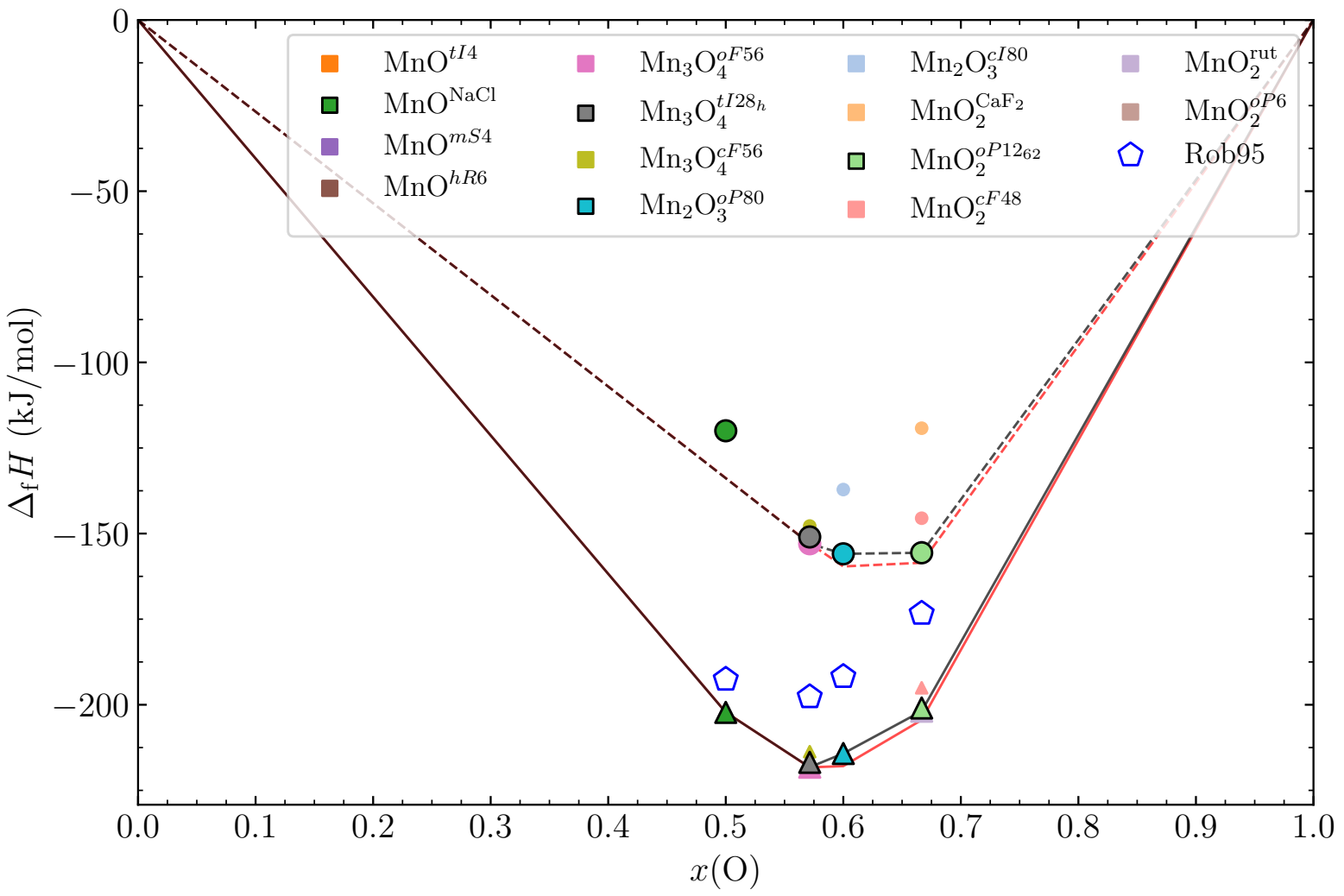


Figure B.8: Ground state of Mn−O system

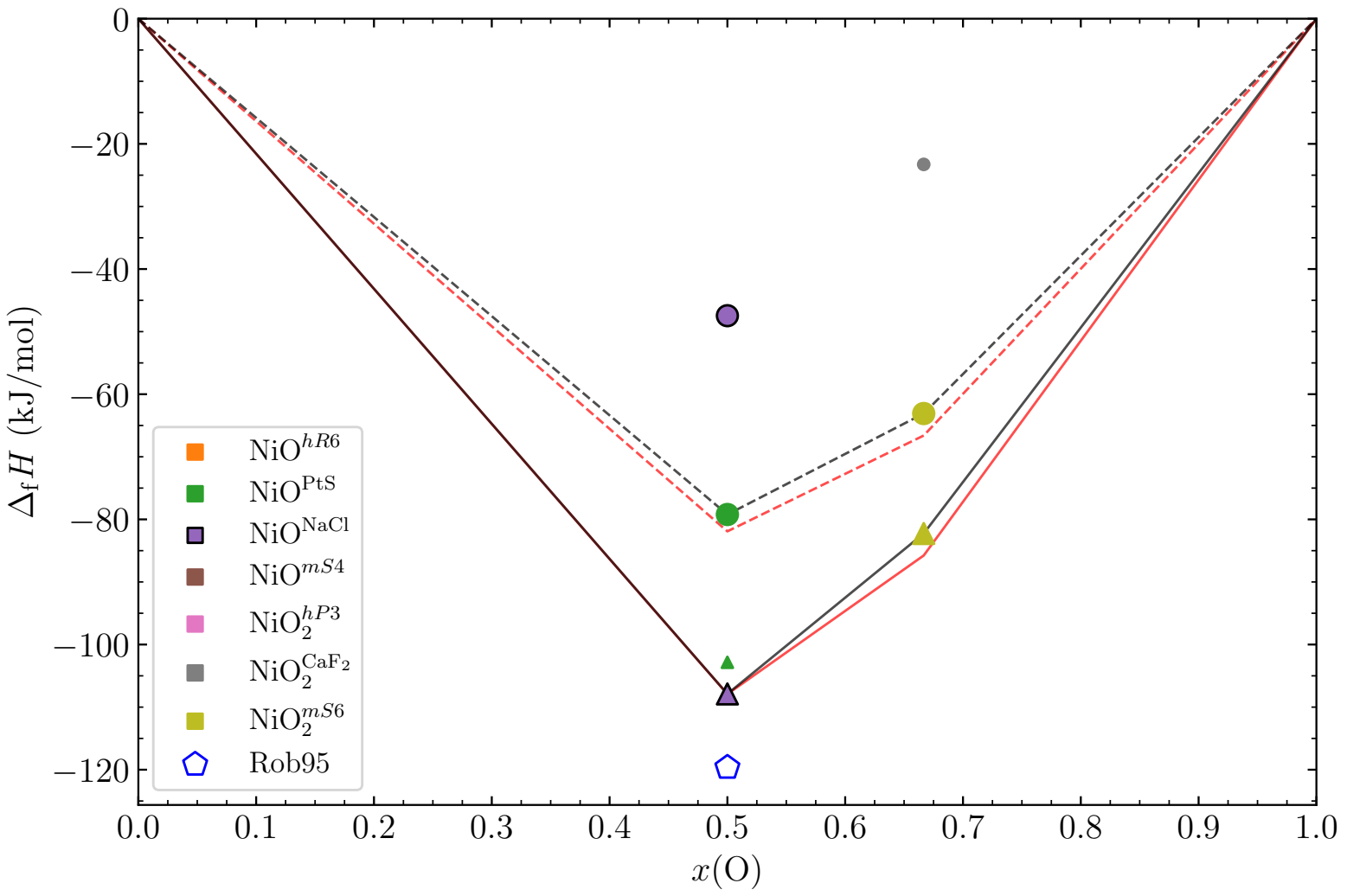


Figure B.9: Ground state of Ni−O system

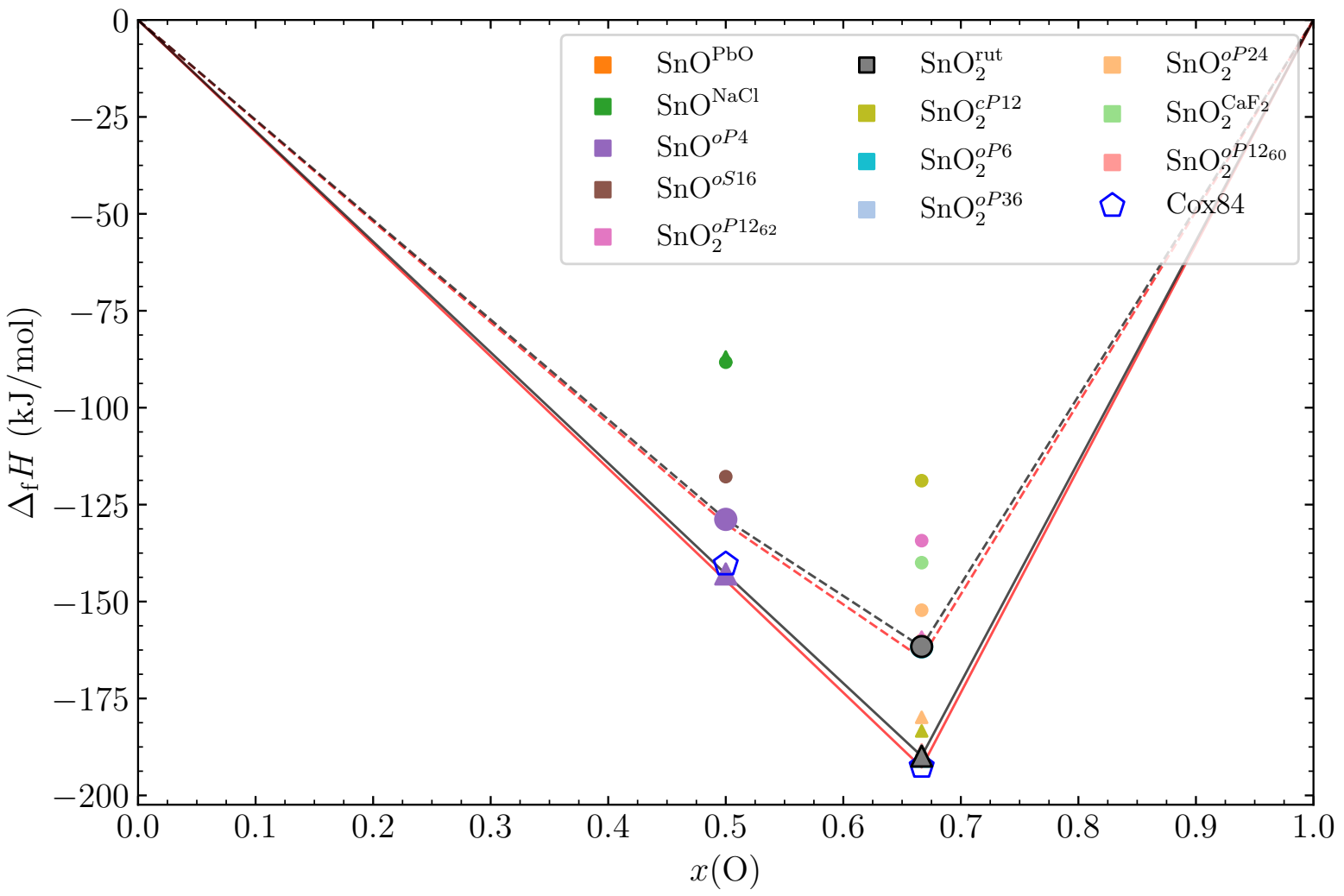


Figure B.10: Ground state of Sn−O system

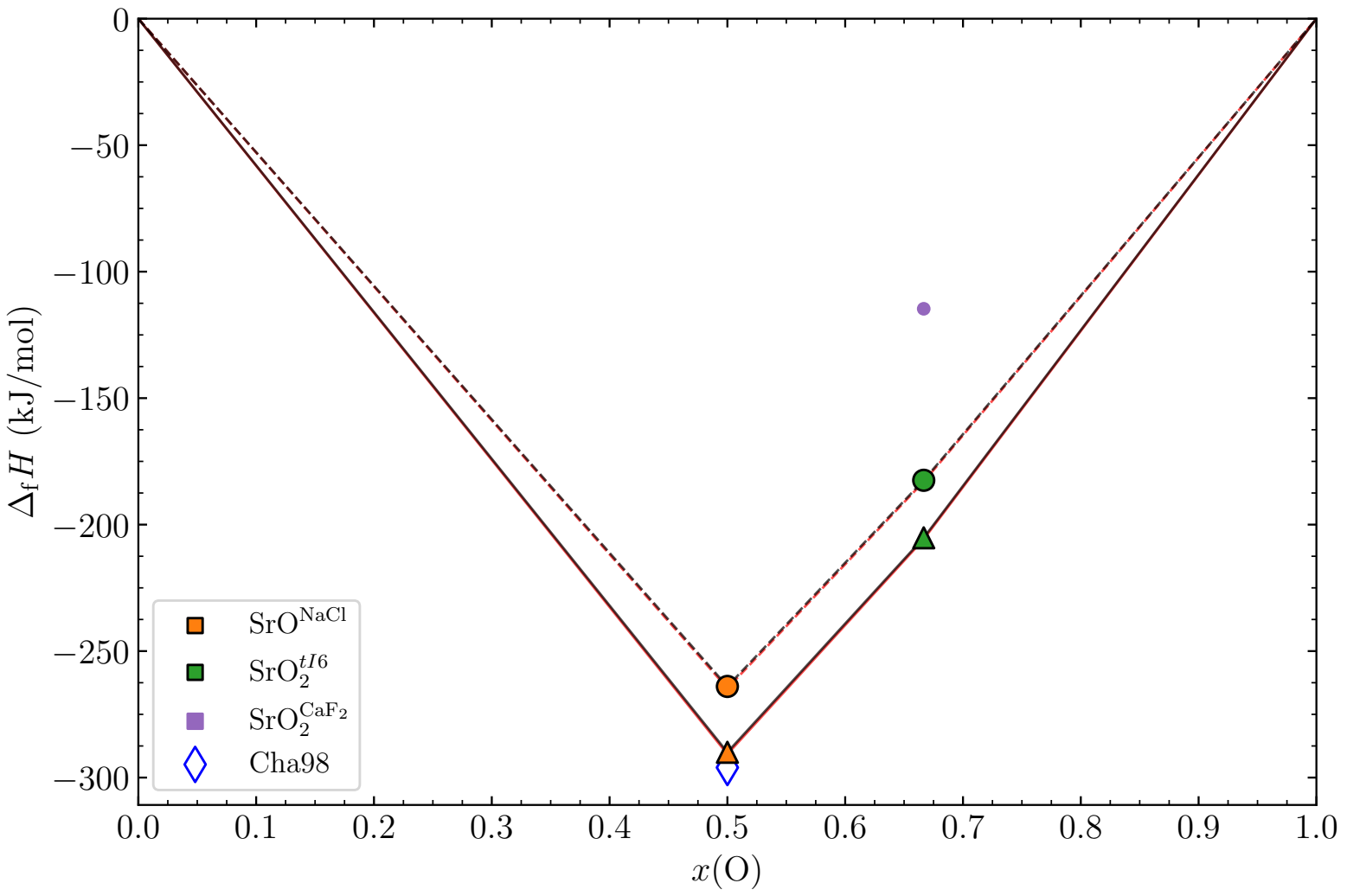


Figure B.11: Ground state of Sr−O system

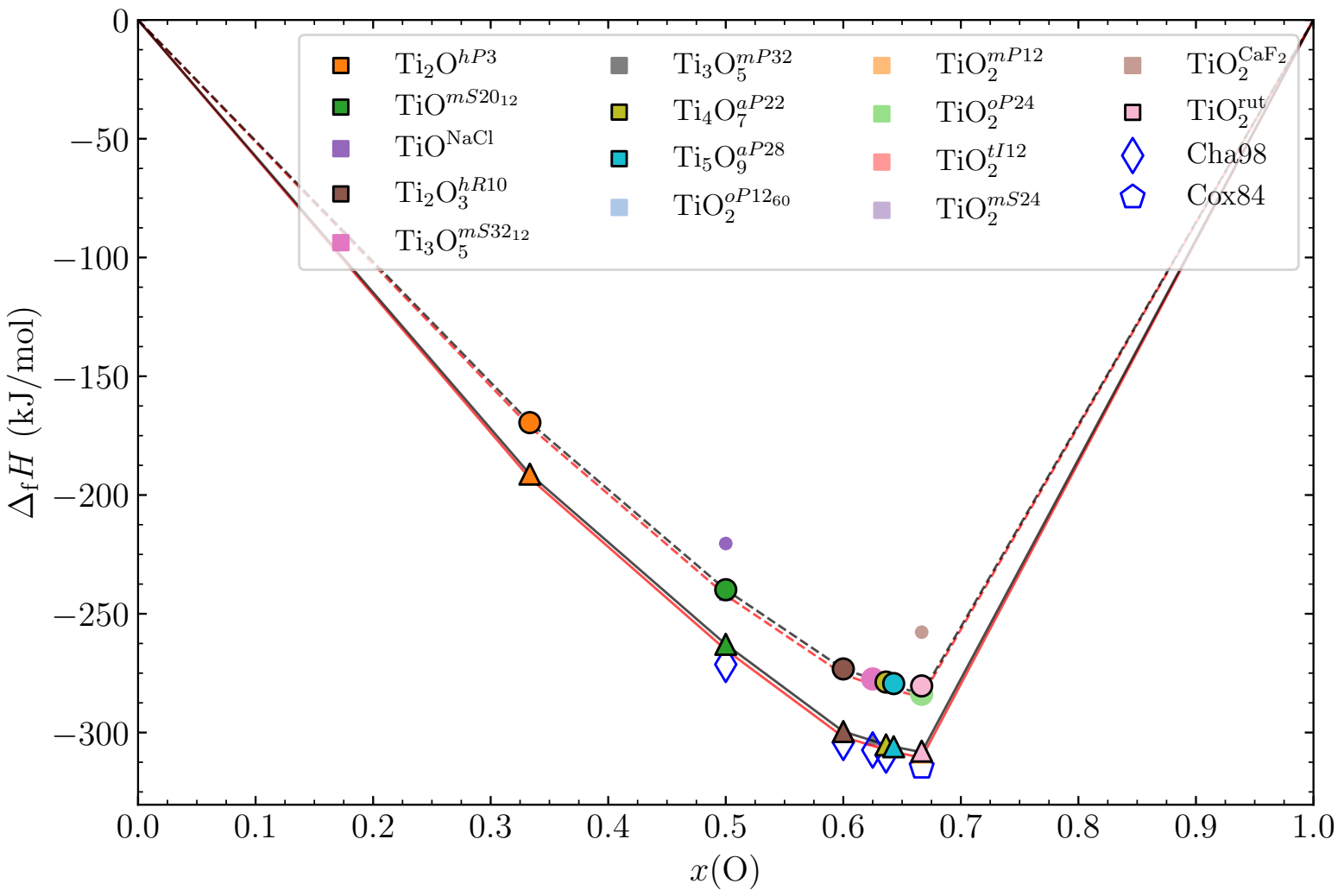


Figure B.12: Ground state of Ti−O system

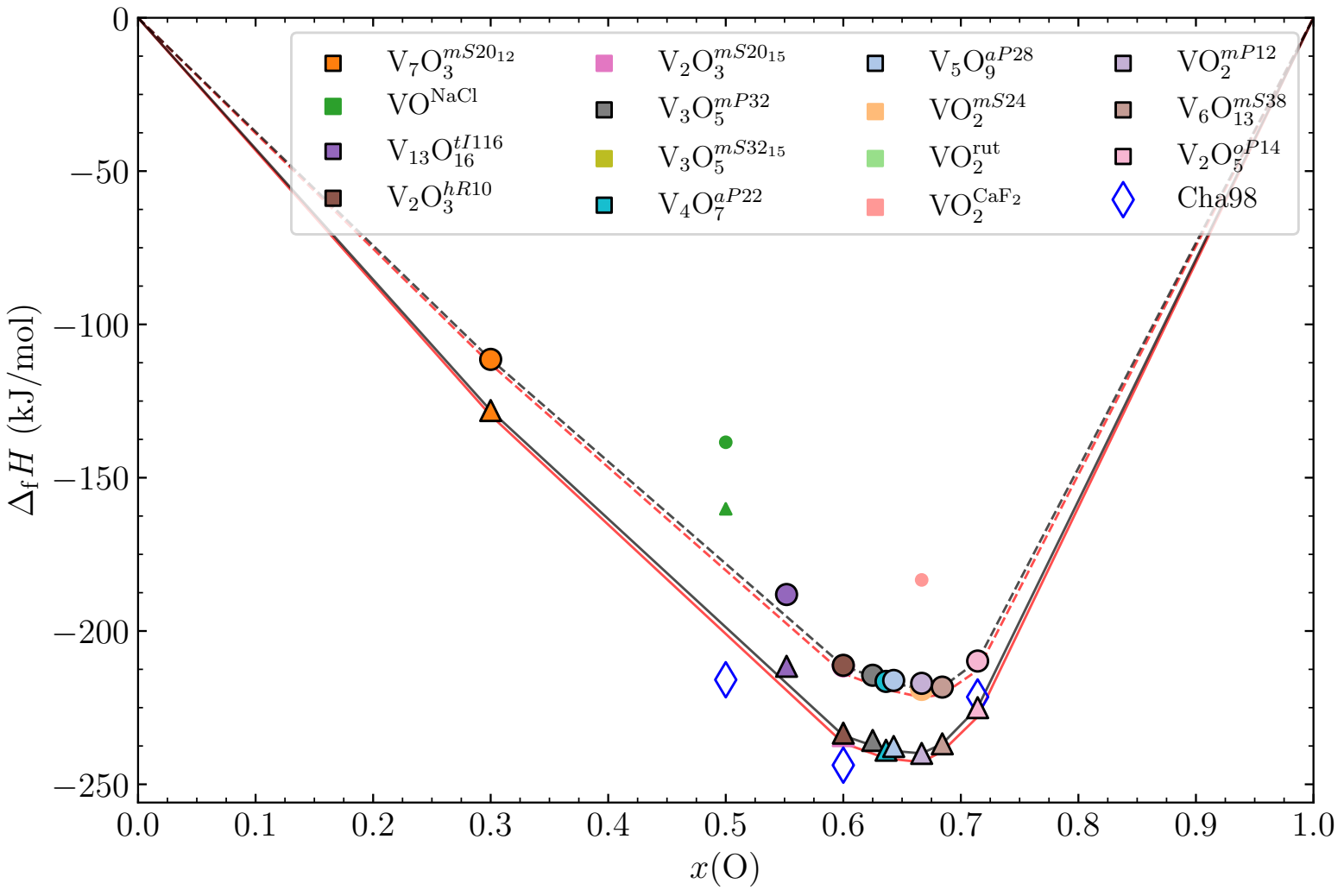


Figure B.13: Ground state of V−O system

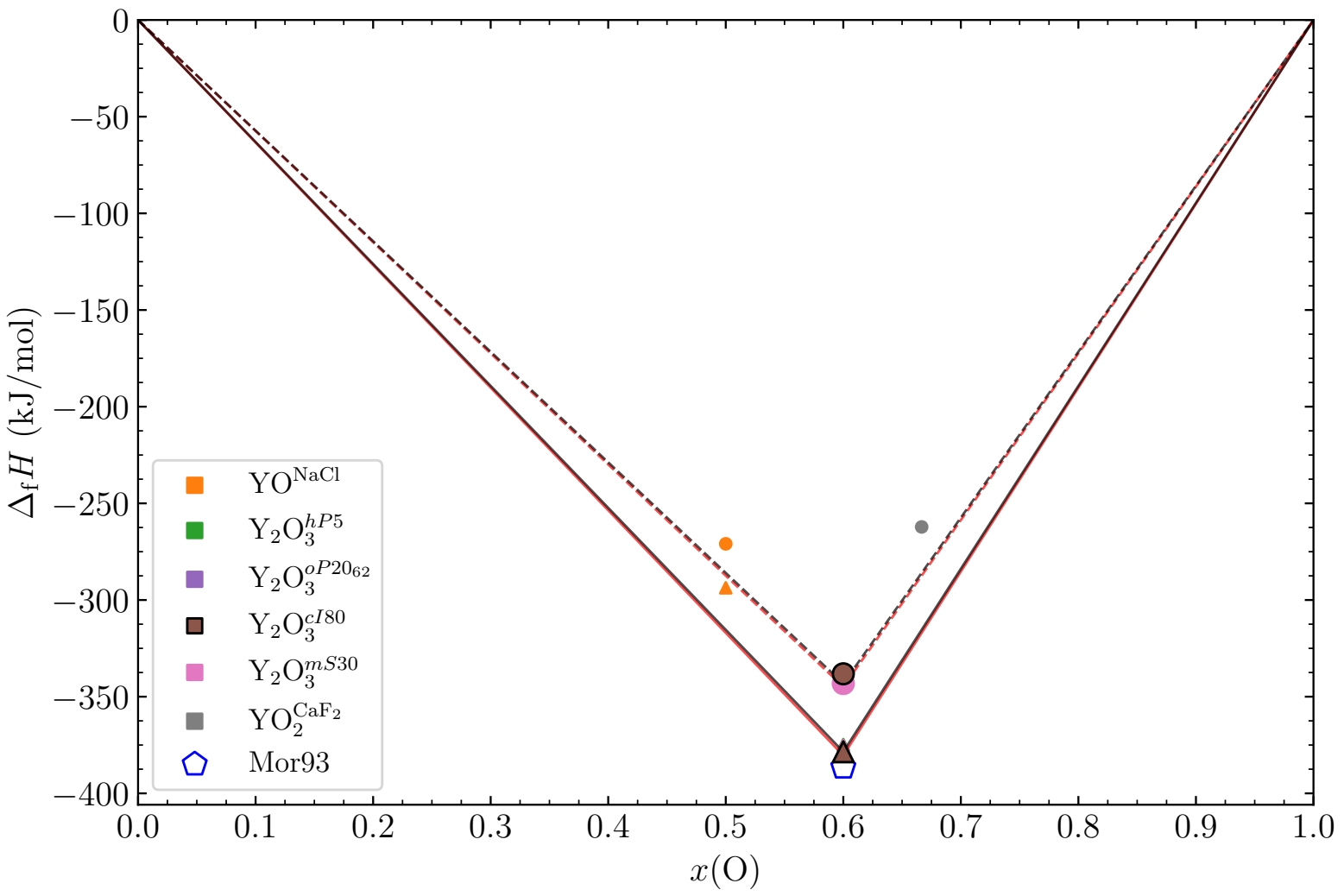


Figure B.14: Ground state of Y−O system

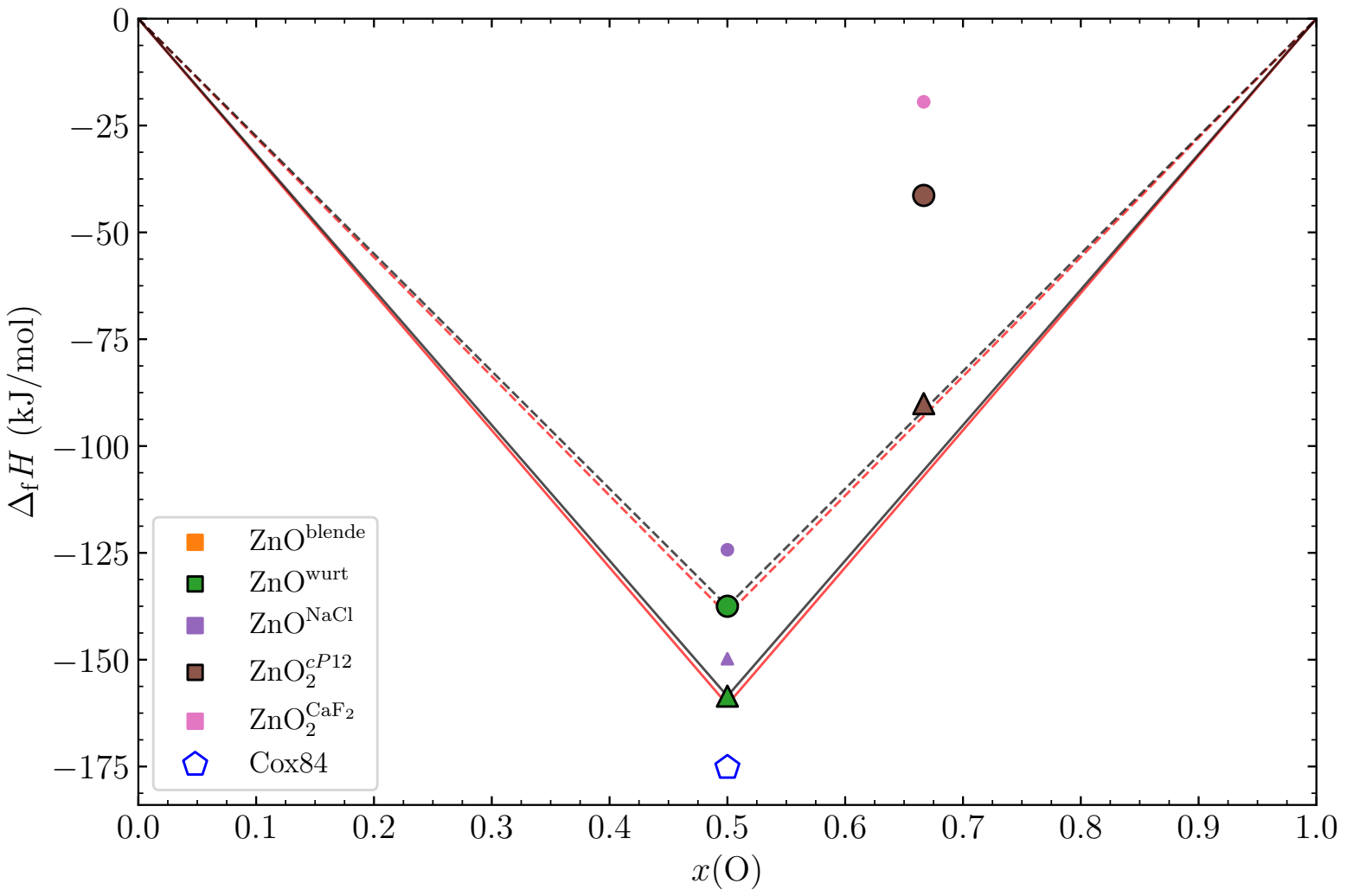


Figure B.15: Ground state of Zn−O system

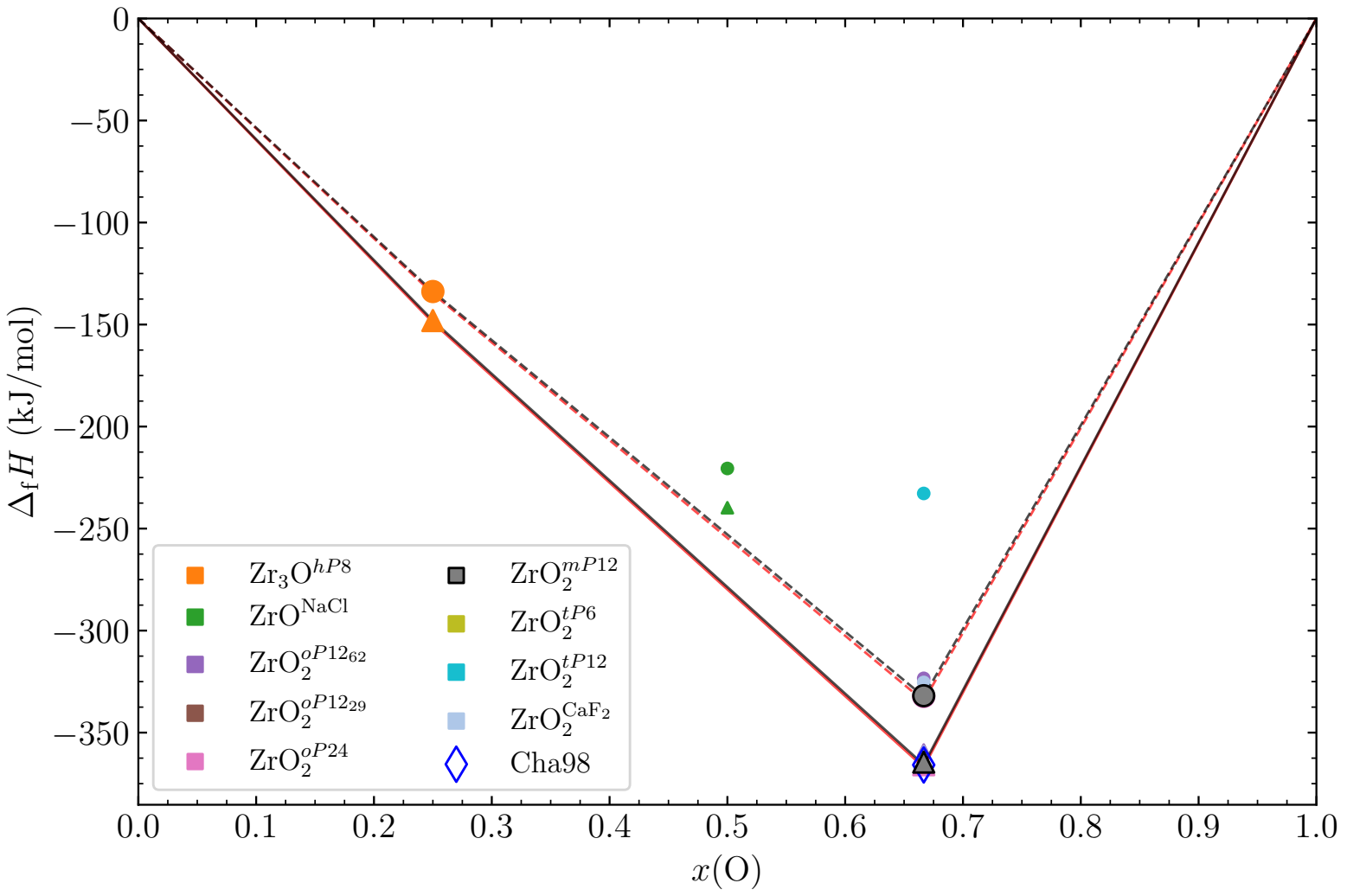


Figure B.16: Ground state of Zr−O system

## Appendix C. Details of ternary $M_1 - M_2$−O ordered compounds

The enthalpy of all ordered ternary compounds is given in Tab. C.1. We observe that compounds can be grouped into four main structural classes:

- perovskites and derivatives (Fig. C.17a),
- structures in the $Fd\bar{3}m$ space group (Fig. C.17b),
- rhombohedral structures (Fig. C.17c),
- $tI14$ phase (Fig. C.17d).

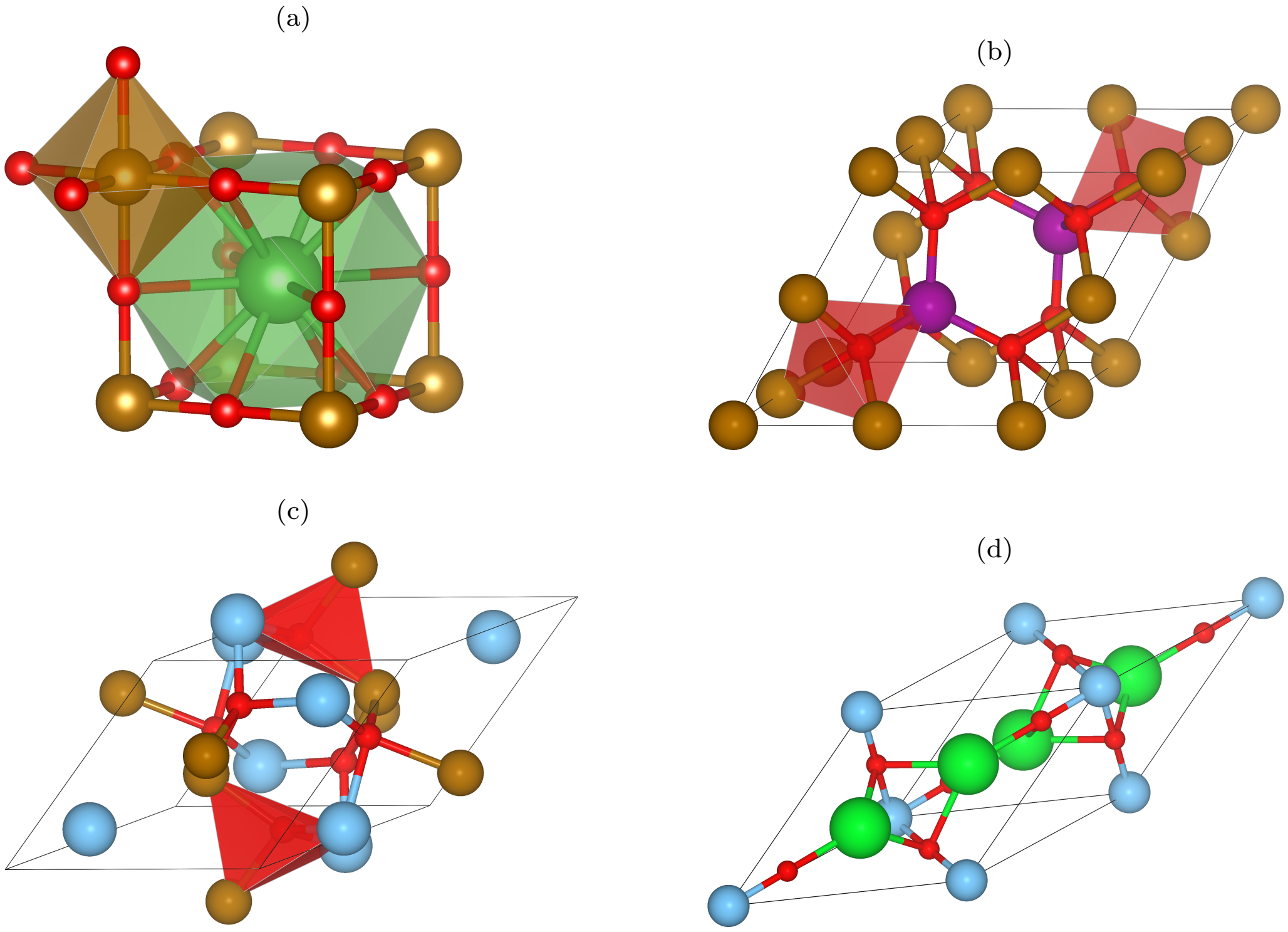


Figure C.17: Primitive cells of the four main classes of ternary oxides in the study.

| Structure | Space Group | Formula | PBE $\Delta_f H$ | PBE $\Delta_{hb} H$ | r$^2$SCAN $\Delta_f H$ | r$^2$SCAN $\Delta_{hb} H$ |
|---|---|---|---|---|---|---|
| $cP5$ | $Pm\bar{3}m$ | $HfSrO_3$ | −326.48 | −14.65 | −358.80 | −16.96 |
| | | $SnSrO_3$ | −213.73 | −11.12 | −245.11 | −15.19 |
| | | $CaTiO_3$ | −293.31 | −8.13 | −321.42 | −11.76 |
| | | $SrZrO_3$ | −316.48 | −11.51 | −347.05 | −11.62 |
| $tP5$ | $P4mm$ | $MnSrO_3$ | −216.02 | −17.06 | −257.93 | −20.81 |
| $oP20_{62}$ | $Pnma$ | $CeVO_3$ | −279.13 | −5.82 | −316.72 | −18.36 |
| | | $CaZrO_3$ | −322.42 | −8.20 | −351.45 | −7.32 |
| $cF56$ | $Fd\bar{3}m$ | $Co_2TiO_4$ | −157.25 | **8.76** | −193.32 | **0.86** |
| | | $Fe_2MnO_4$ | −111.06 | **11.00** | −181.25 | −9.51 |
| $cF88$ | | $Ce_2Zr_2O_7$ | −332.16 | −5.33 | −365.29 | −8.62 |
| | | $Hf_2Y_2O_7$ | −346.71 | −3.24 | −380.19 | −2.83 |
| $aP22$ | | $Ca_2V_2O_7$ | −266.39 | −28.49 | −286.64 | −30.09 |
| | | $Mg_2V_2O_7$ | −243.46 | −15.64 | −263.81 | −15.75 |
| $E2_2$ | $R\bar{3}$ | $FeTiO_3$ | −210.26 | −2.75 | −242.47 | −5.53 |
| | | $CoTiO_3$ | −196.43 | **4.95** | −235.27 | −6.87 |
| | | $CaSnO_3$ | −223.40 | −11.54 | −248.49 | −9.87 |
| | | $MgSnO_3$ | −205.21 | −4.44 | −232.37 | −3.08 |
| | | $TiZnO_3$ | −225.30 | **0.04** | −251.98 | −3.61 |
| | | $MnZnO_3$ | −152.56 | −4.20 | −191.21 | −6.68 |
| | | $MgTiO_3$ | −281.17 | −7.08 | −308.96 | −8.63 |
| | | $MnTiO_3$ | −228.86 | −4.98 | −277.54 | −11.69 |
| | | $MnSnO_3$ | −150.27 | 2.97 | −198.76 | −1.92 |
| | | $MgMnO_3$ | −206.47 | −9.36 | −246.51 | −10.01 |
| $hR10$ | $R3c$ | $CaHfO_3$ | −330.02 | −8.95 | −359.57 | −9.01 |
| | | $SrTiO_3$ | −299.14 | −23.22 | −329.24 | −28.28 |
| | | $SnZnO_3$ | −150.61 | **1.41** | −176.97 | **0.36** |
| | | $NiTiO_3$ | −195.11 | **6.92** | −233.99 | −5.87 |
| | | $CaMnO_3$ | −219.85 | −11.64 | −259.77 | −13.95 |
| $hR24$ | $R\bar{3}c$ | $Ca_2Hf_7O_{16}$ | −339.80 | −5.13 | −371.81 | −5.74 |

Table C.1: Formation enthalpy ($\Delta_f H$) and deviation from binary ground-state ($\Delta_{hb} H$) in kJ/mol, of ternary oxides compute with PBE and r$^2$SCAN functionnal.

| Structure | Space Group | Formula | PBE | | r$^2$SCAN | |
|---|---|---|---|---|---|---|
| | | | $\Delta_f H$ | $\Delta_{hb} H$ | $\Delta_f H$ | $\Delta_{hb} H$ |
| $tI14$ | $I4/mmm$ | $Ca_2TiO_4$ | −293.17 | −7.44 | −319.91 | −9.65 |
| | | $Ca_2MnO_4$ | −240.61 | −9.86 | −277.12 | −12.45 |
| | | $Sr_2TiO_4$ | −292.14 | −19.63 | −321.29 | −23.46 |
| | | $MnSr_2O_4$ | −236.08 | −18.55 | −274.27 | −22.04 |
| | | $CuY_2O_4$ | −261.29 | −0.04 | −289.69 | **0.65** |
| $mS18$ | $P2/c$ | $MgV_2O_6$ | −233.02 | −12.21 | −250.59 | −11.49 |
| | | $CaV_2O_6$ | −248.76 | −21.79 | −265.53 | −21.25 |
| $mS224$ | $C2/c$ | $CaZr_4O_9$ | −329.93 | −4.09 | −361.36 | −3.34 |
| $oS32$ | $Cmcm$ | $MgTi_2O_5$ | −283.42 | −5.65 | −309.52 | −6.21 |
| $oS38$ | $Cmmm$ | $Sr_4Fe_4O_{11}$ | −194.27 | −23.40 | −225.00 | −26.82 |
| $oP14$ | $Pbam$ | $Sr_2CeO_4$ | −300.13 | −6.02 | −326.52 | −5.15 |

Table C.1: Formation enthalpy ($\Delta_f H$) and deviation from binary ground-state ($\Delta_{hb} H$) in kJ/mol, of ternary oxides compute with PBE and r$^2$SCAN functionnal.(continued)

## Appendix D. Training database analysis

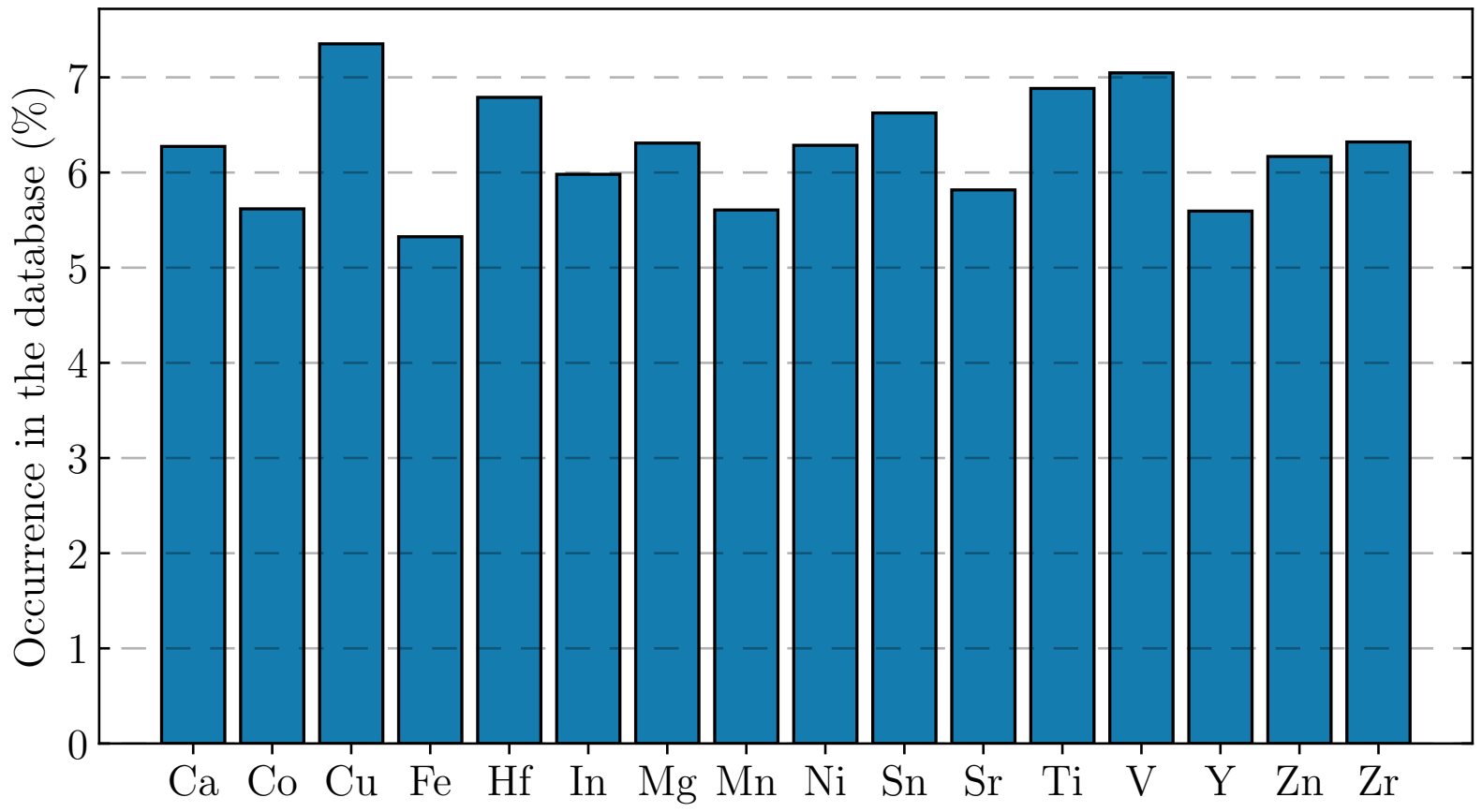


Figure D.18: Occurrence of chemical cationic elements in the ESO training database.

<table>
<tr><th>Compound</th><th>Number</th><th>Set</th><th>%</th></tr>
<tr><td>Ordered</td><td>16</td><td rowspan="6">Train</td><td rowspan="6">82</td></tr>
<tr><td>Binary</td><td>120</td></tr>
<tr><td>Ternary</td><td>27</td></tr>
<tr><td rowspan="4">416 Quinary</td><td>104</td></tr>
<tr><td>104</td></tr>
<tr><td>104</td></tr>
<tr><td>104</td><td>Test</td><td>18</td></tr>
</table>

Table D.2: Overview of the training database composition.

## Appendix E. Performance of machine learning models

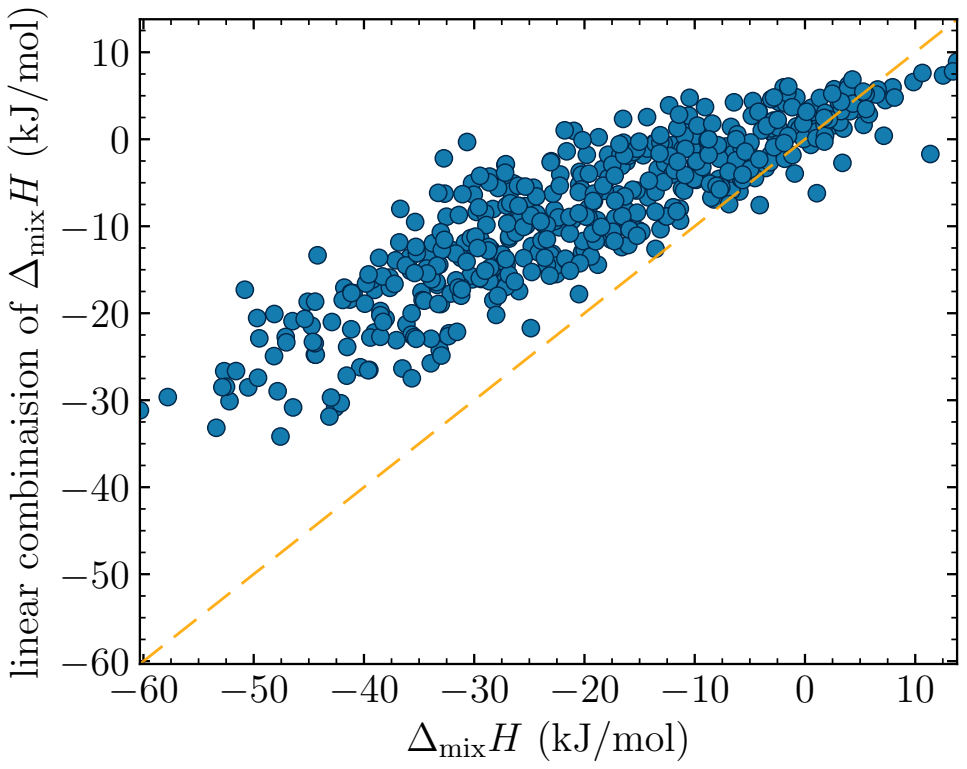


Figure E.19: Mixing enthalpy of quinary systems, extrapolated from a linear combination of binary systems.

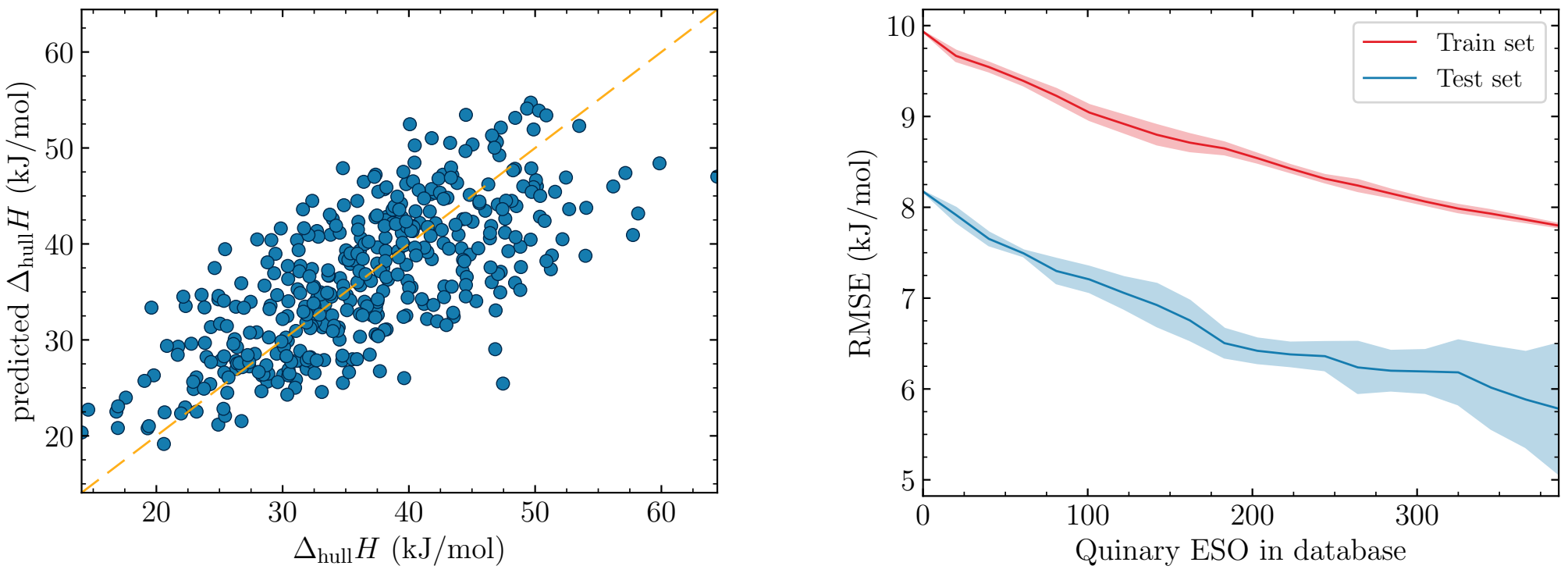


Figure E.20: Prediction and learning curves of linear model.

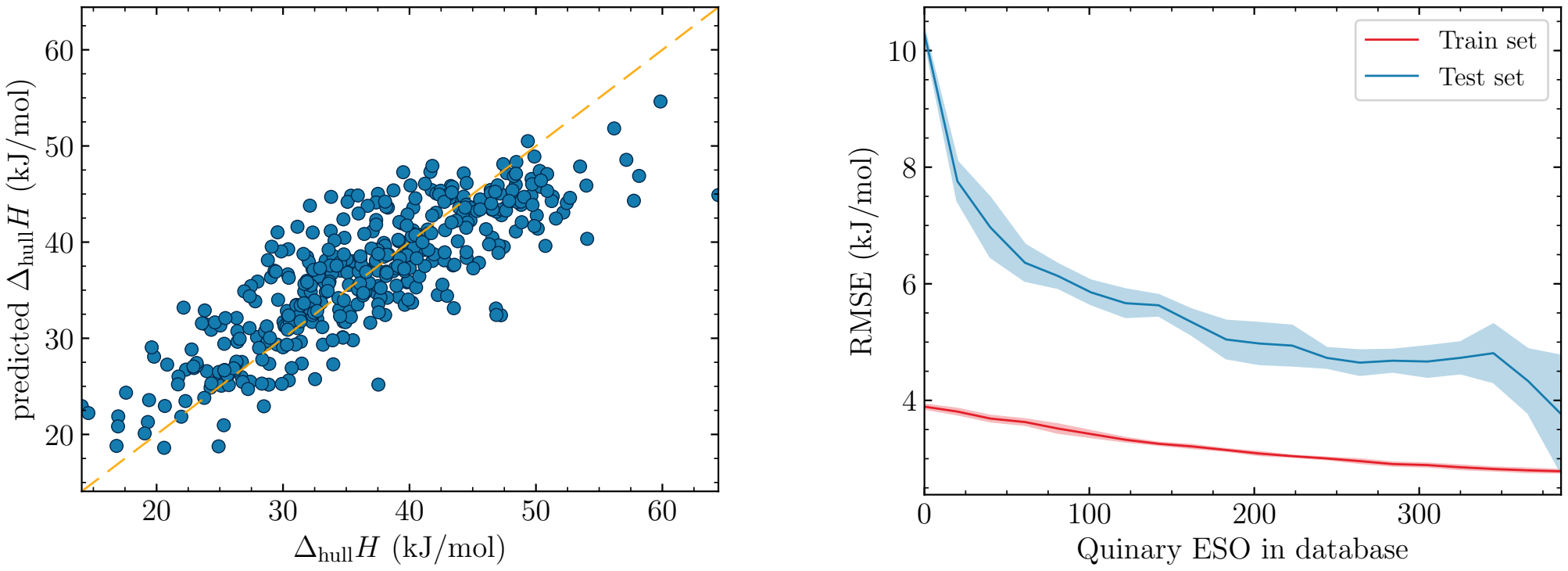


Figure E.21: Prediction and learning curves of Random Forest model.

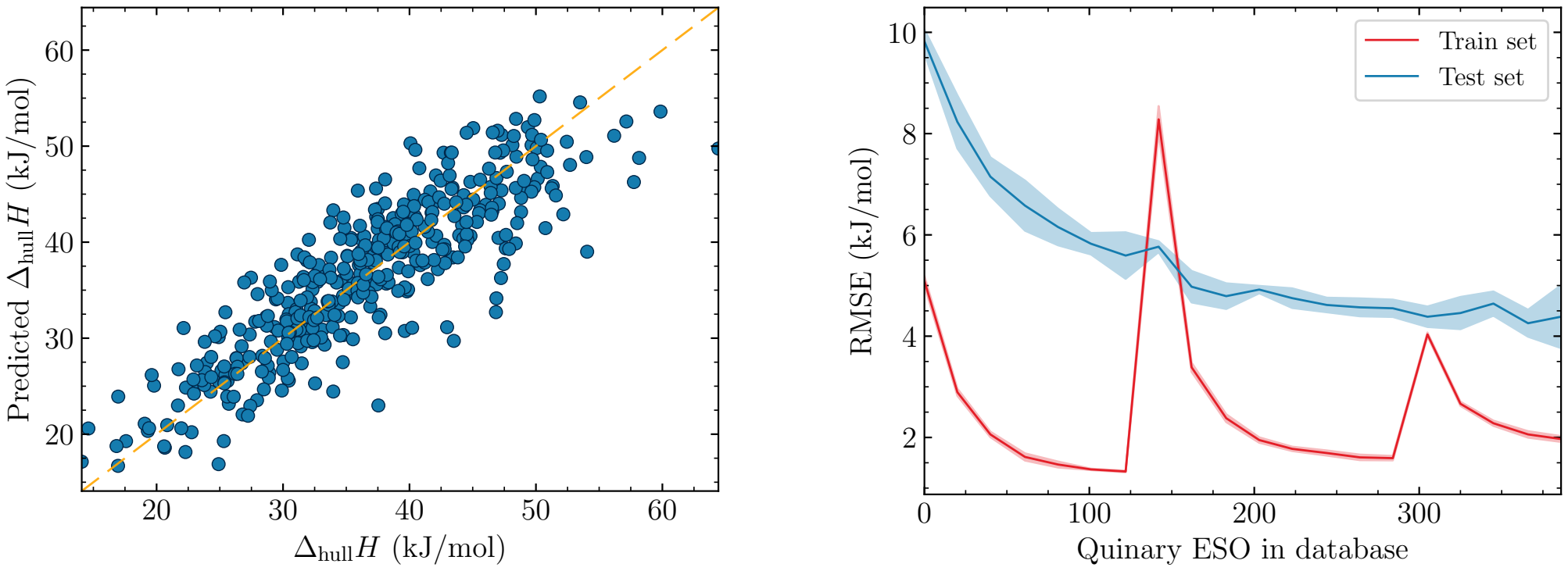


Figure E.22: Prediction and learning curves of Neural Network model.

## Appendix F. Prediction of the whole combination of HEO

Scripts used for postprocessing will be deposited together with the data (DFT-calculated training database and prediction of the whole combination of HEO) in the following URL after publication:

`https://github.com/link-lab3629`

The most promising candidates sorted by stabilization temperature (increasing order) are given in the table of the following page.

| Compound | $T_{\text{stab}}$ | Compound | $T_{\text{stab}}$ | Compound | $T_{\text{stab}}$ |
|---|---|---|---|---|---|
| (Ca,Cu,Fe,Mg,Zn)O | 1,676 | (Ca,Cu,In,Sr,Zn)O | 1,750 | (Ca,Cu,Ni,Sr,Zn)O | 1,868 |
| (In,Sn,Sr,V,Zn)O | 2,080 | (Ca,Cu,Fe,Ni,Zn)O | 2,151 | (Ca,Cu,In,Mg,Sr)O | 2,153 |
| (Ca,Co,Cu,Fe,Zn)O | 2,183 | (Ca,Co,Cu,Mg,Sr)O | 2,214 | (Ca,Co,Cu,Sr,Zn)O | 2,281 |
| (Ca,In,Mg,Sr,Zn)O | 2,302 | (Ca,Co,Cu,Fe,Mg)O | 2,307 | (Ca,In,Sn,Sr,Zn)O | 2,323 |
| (Ca,Co,Cu,Mg,Zn)O | 2,395 | (Ca,Mg,Mn,Sr,Zn)O | 2,408 | (In,Mn,Sn,Sr,Zn)O | 2,413 |
| (Ca,In,Mg,Mn,Zn)O | 2,415 | (Mn,Sn,Sr,V,Zn)O | 2,427 | (Cu,In,Mg,Sr,Zn)O | 2,455 |
| (Ca,Cu,Mg,Ni,Zn)O | 2,467 | (Ca,In,Mn,Sr,Zn)O | 2,496 | (Ca,Co,Cu,Mg,Ni)O | 2,515 |
| (Ca,In,Sn,Sr,Y)O | 2,523 | (Co,Cu,Fe,Mn,Ni)O | 2,535 | (Ca,Cu,Fe,Sr,Zn)O | 2,551 |
| (Ca,Co,Cu,Ni,Sr)O | 2,578 | (Ca,Cu,Mg,Sr,Zn)O | 2,581 | (Co,Cu,Mg,Sr,Zn)O | 2,585 |
| (Ca,In,Mg,Sn,Sr)O | 2,609 | (Ca,Mg,Ni,Sr,Zn)O | 2,619 | (Co,Cu,Fe,Mg,Ni)O | 2,627 |
| (Ca,Mg,Sn,Sr,Zn)O | 2,674 | (Mg,Mn,Sn,Sr,Zn)O | 2,696 | (Ca,Co,Ni,Sr,Zn)O | 2,700 |
| (Ca,Cu,Mg,Ni,Sr)O | 2,700 | (Ca,In,Mn,Sn,Sr)O | 2,711 | (Ca,Co,Mg,Sr,Zn)O | 2,745 |
| (Co,Cu,Mg,Ni,Zn)O | 2,745 | (Ca,Fe,Mg,Mn,Zn)O | 2,747 | (In,Mg,Mn,Sn,Sr)O | 2,805 |
| (In,Mg,Sn,Sr,Zn)O | 2,836 | (Ca,Mn,Sn,Sr,Zn)O | 2,837 | (Cu,In,Mg,Ni,Sr)O | 2,848 |
| (Ca,Fe,In,Sr,Zn)O | 2,863 | (Ca,Co,Cu,In,Sr)O | 2,880 | (Ca,Co,Cu,Ni,Zn)O | 2,886 |
| (Ca,Cu,Fe,Mg,Sr)O | 2,900 | (Ca,Mg,Mn,Sn,Sr)O | 2,901 | (Fe,In,Sr,V,Zn)O | 2,931 |
| (Ca,Fe,Ni,Sr,Zn)O | 2,991 | (Ca,Co,Cu,In,Ni)O | 3,044 | (Cu,In,Ni,Sr,Zn)O | 3,048 |
| (Co,Cu,Fe,Ni,Zn)O | 3,078 | (Ca,Fe,Mn,Ni,Zn)O | 3,086 | (Ca,In,Mn,Sn,Zn)O | 3,114 |
| (Ca,Sn,Sr,Y,Zn)O | 3,160 | (Co,Fe,Mg,Mn,Ni)O | 3,177 | (Co,Mg,Ni,Sr,Zn)O | 3,203 |
| (Cu,Mg,Ni,Sr,Zn)O | 3,205 | (Ca,In,Sr,V,Zn)O | 3,241 | (Co,Cu,Fe,Sr,Y)O | 3,247 |
| (Ca,Co,Fe,Sr,Zn)O | 3,263 | (Ca,Co,Cu,Mn,Ni)O | 3,281 | (Fe,Mn,Sn,Sr,V)O | 3,308 |
| (Ca,Cu,Fe,Mg,Ni)O | 3,314 | (Cu,Fe,Mn,Sr,Zn)O | 3,330 | (Cu,Fe,In,Sr,Zn)O | 3,331 |
| (Cu,In,Sr,V,Y)O | 3,336 | (Ca,Cu,Fe,Ni,Sr)O | 3,337 | (Co,Cu,Mg,Ni,Sr)O | 3,359 |
| (Cu,In,Mn,Sn,Sr)O | 3,403 | (Ca,In,Mg,Sn,Y)O | 3,424 | (Co,Cu,Fe,Sr,Zn)O | 3,429 |

Table F.3: Quinary ESO with the lowest stabilization temperatures ($T_{\text{stab}}$, in K), calculated within the GGA framework. Known ESOs are highlighted in red, mixtures shown in blue melted before ESO stabilization occurred, whereas those shown in green partially formed ESO phases.


\end{document}